\documentclass[12pt]{article}
\usepackage{epsfig,amssymb,amsmath,latexsym,bbm}
\title{Scaling diagram for the localization length at a band edge}

\author{Christian Sadel, Hermann Schulz-Baldes
\\
\\
{\small Mathematisches Institut, Universit\"at
Erlangen-N\"urnberg, Germany}
}

\date{ }

\newtheorem{theo}{Theorem}
\newtheorem{defini}{Definition}
\newtheorem{proposi}{Proposition}
\newtheorem{lemma}{Lemma}

\newcommand{\NN}{{\mathbb N}}
\newcommand{\RR}{{\mathbb R}}

\newcommand{\ZZ}{{\mathbb Z}}

\newcommand{\PP}{{\bf P}}
\newcommand{\pp}{{\bf p}}
\newcommand{\EE}{{\bf E}}

\newcommand{\Ss}{{\cal S}}
\newcommand{\Oo}{{\cal O}}
\newcommand{\Tr}{\mbox{\rm Tr}}
\newcommand{\Tt}{{\cal T}}
\newcommand{\Rr}{{\cal R}}
\newcommand{\Nn}{{\cal N}}

\newcommand{\Ll}{{\cal L}}

\newcommand{\Kk}{{\mathcal K}}

\newcommand{\comm}[1]{}
\newcommand{\comment}[1]{}

\begin{document}

\maketitle

\begin{abstract}
A weak-coupling scaling diagram for the
Lyapunov exponent and the integrated density of states near a
band edge of a random Jacobi matrix is obtained. The
analysis is based on the use of a Fokker-Planck operator 
describing the drift-diffusion of the Pr\"ufer phases.
\end{abstract}

\vspace{.9cm}

\section{Main result and short overview}
\label{sec-intro}

This work considers one-dimensional discrete random
Schr\"odinger operators, typically 
given by a periodic background operator and a 
weakly-coupled random potential. The periodic (non-random) 
operator has a band structure. In the
vicinity of a band edge, we present a rigorous pertubation
theory in the coupling constant and the energy for the
Lyapunov exponent and the density of states.
This leads to a new and rich
scaling diagram describing these two self-averaging quantities at
a band edge.

\vspace{.2cm}

The scaling behavior at energies within a band has been well
understood for a long time. For sake of concreteness, let us focus on
the Anderson model $H=\Delta + \lambda V$
given as the sum of the discrete Laplacian $\Delta$ and a disordered potential
$V=\sum_{n\in\ZZ}v_n\;|n\rangle\langle n|$ coupled with a (small) coupling
constant $\lambda\geq 0$. Thouless \cite{Tho} found a perturbative formula for
the Lyapunov exponent (inverse localization length) at an energy
$E=2\,\cos(k)$ in the band spectrum of $\Delta$:
\begin{equation}
\label{eq-Thouless}
\gamma_\lambda(E)
\;=\;
\frac{\EE_\sigma(v_\sigma^2)}{8\,\sin^2(k)}
\;\lambda^2
\;+\;\Oo(\lambda^3)
\;,
\end{equation}
where $v_\sigma$ is one of the identically distributed and centered 
entries of $V$ and $\EE_\sigma$ denotes the expectation value. A
rigorous proof was provided by Pastur and Figotin using the 
Pr\"ufer phases, also called the 
Dyson-Schmidt variables \cite{PF}. This amounts to justifying the
so-called random phase approximation, stating that the Pr\"ufer phases are
distributed according to the Lebesgue measure on the unit circle. A further
generalization was given in \cite{JSS}. One feature of \eqref{eq-Thouless} are
the singularities at the
band edges  $k=0,\pi$ of the unperturbed operator $\Delta$. 
In fact, the control on the error terms in 
\eqref{eq-Thouless} breaks down at those energies, 
as does the random phase approximation. 
The latter also happens at
the band center, namely for $k=\frac{\pi}{2},\frac{3\pi}{2}$; 
this leads to anomalies in the perturbative formula \eqref{eq-Thouless}
first
found by Kappus and Wegner \cite{KW}, and consecutively analyzed by several
authors \cite{DG,BK,CK,SVW}. A perturbative formula as \eqref{eq-Thouless} for
the band center
with a control on the error terms was proven only recently \cite{S}.

\vspace{.2cm}

This paper considers the band edges and shows how the $\lambda^2$ is modified,
leading to the new scaling diagram near a band edge. Parts of this diagram were
already given by Derrida and Gardner \cite{DG}. 
These authors actually found the correct scaling in the parabolic regime of
Theorem~\ref{theo-scaling} below, but the wrong prefactor ({\it cf.}
the comment at the end of Section~\ref{sec-bandedge}). 
Moreover, they
could not give a better justification of
their scaling Ansatz than that it leads to a differential equation they could
solve. Our more conceptual approach shows why the scaling is natural in the
situation considered in \cite{DG}. It exhibits a far richer scaling
behavior near a band edge and also allows to rigorously control the
higher order corrections. Moreover, if one wants to give a
perturbative proof of uniform positivity of the Lyapunov exponent
in an energy interval around a band edge, all of the 
scaling behaviors considered below are needed.

\vspace{.2cm}

Let us now describe the main result in more detail.
We consider a one-parameter family of 
random Jacobi matrices $(H_{\lambda,\omega})_{\lambda\geq 0}$ given as the 
sum of an
$L$-periodic background operator $H_0=H_{0,\omega}$ and a random
perturbation $H_{\lambda,\omega}-H_0$ which is linear in the (small)
coupling constant $\lambda$. More precisely, for every fixed 
configuration $\omega$,
the operator $H_{\lambda,\omega}$ acts on $\psi\in\ell^2(\ZZ)$ as
\begin{equation}
\label{eq-Jacobi}
H_{\lambda,\omega}|n\rangle
\;=\;
t_{\lambda,\omega}(n+1)|n+1\rangle
+v_{\lambda,\omega}(n)|n\rangle+
t_{\lambda,\omega}(n)|n-1\rangle
\;,
\qquad
n\in\ZZ\;,
\end{equation}
where the coefficients $t_{\lambda,\omega}(n)>0$ and 
$v_{\lambda,\omega}(n)\in\RR$ are constructed as described in the
following: let
$(\hat{t}_1,\ldots,\hat{t}_L,\hat{v}_1,\ldots,\hat{v}_L)$
be given real constants with $\hat{t}_l>0$ for $l=1,\ldots,L$;
set $\Sigma=[-1,1]^{2L}$ so that each
$\sigma\in\Sigma$ is of the form
$\sigma=(\tilde{t}_1(\sigma),\ldots,
\tilde{t}_L(\sigma),\tilde{v}_1(\sigma),\ldots,\tilde{v}_L(\sigma))$;
then $\Omega=\Sigma^\ZZ\times\{1,\ldots,L\}$ is the configuration space and
to each $\omega=((\sigma_m)_{m\in\ZZ},k)\in\Omega$ there are 
associated sequences
$$
t_{\lambda,\omega}(k-1+Lm+l)
\:=\;
\hat{t}_l+\lambda\,\tilde{t}_l(\sigma_m)
\:,
\qquad
v_{\lambda,\omega}(k-1+Lm+l)
\:=\;
\hat{v}_l+\lambda\,\tilde{v}_l(\sigma_m)
\:,
$$
which for $\lambda$ sufficiently small satisfy $t_{\lambda,\omega}(n)>0$;
these sequences define $H_{\lambda,\omega}$ by \eqref{eq-Jacobi}. In
order to make $(H_{\lambda,\omega})_{\lambda\geq 0}$ into a
family of random operators, we equip $\Omega$ with a probability measure 
$\PP={\bf p}^\ZZ\times \frac{1}{L}\sum_{l=1}^L \delta_l$
where ${\bf p}$ is a probability measure on $\Sigma$. Expectation
values w.r.t. $\PP$ and ${\bf p}$ will be denoted by $\EE$ and
$\EE_\sigma$. We suppose that $\EE_\sigma(\tilde{t}_l(\sigma))
=\EE_\sigma(\tilde{v}_l(\sigma))=0$.

\vspace{.2cm}

For every fixed energy $E\in\RR$ and coupling parameter $\lambda$, 
there are two self-averaging
quantities of interest, 
namely the integrated density of states (IDS) $\Nn_\lambda(E)$ and
the Lyapunov exponent (or inverse localization length)
$\gamma_\lambda(E)$. The definitions will be recalled
in Section~\ref{sec-dynamics} below.
The periodic operator $H_0=H_{0,\omega}$ has a
band structure and we are interested in the scaling of the IDS and
Lyapunov exponent at one of its band edges $E_b$ (band touching excluded). In
order to state the precise result, we need to introduce two quantities.
Let $\Tt^E_{\lambda,\sigma}$ be the random transfer matrix at energy $E$
over a unit cell of
length $L$ (see Section~\ref{sec-normal} for the explicit formula). Then set
$$
x
\;=\;
\left.
\partial_E\,\Tr(\Tt^E_{0,\sigma})
\right|_{E=E_b}
\;,
\qquad
x_\sigma
\;=\;
\partial_\lambda\,
\Tr(\Tt^{E_b}_{\lambda,\sigma})
\,|_{\lambda=0}
\;.
$$

\begin{theo}
\label{theo-scaling}
Let $E_b$ be a band edge {\rm (}band touching excluded{\rm )}
of the periodic background operator
$H_0=H_{0,\omega}$ of a family of random
Jacobi matrices $(H_{\lambda,\omega})_{\lambda\geq 0}$.
The perturbation is supposed to be non-trivial in
the sense that $x_\sigma$ does not vanish ${\bf p}$-almost surely. 
Then the scaling near the band edge is 
\begin{equation}
\label{eq-scaling}
\Nn_\lambda(E_b+\epsilon\,\lambda^\eta)
\;=\;
\Nn_0(E_b)\,+\,
A\,\lambda^\alpha
\,+\,\Oo(\lambda^{\alpha+\delta})
\;,
\qquad
\gamma_\lambda(E_b+\epsilon\,\lambda^\eta)
\;=\;
B\,\lambda^\beta
\,+\,\Oo(\lambda^{\beta+\delta})
\;,
\end{equation}
where $A,B,\alpha,\beta$ and $\delta>0$ depend on
$\epsilon\in\RR$ and $\eta>0$ as described in the following and
resumed in {\rm Figure~1}.

\vspace{.2cm}

\noindent {\rm (i) (Elliptic regime)} Let $\eta<\frac{4}{3}$ and let
the sign of $\epsilon\neq 0$ be
such that $E_b+\epsilon\,\lambda^\eta$ is inside the band of
$H_0$. For the case of the Lyapunov exponent, we also suppose 
$\eta>\frac{4}{5}$. Then
$$
\alpha
\;=\;
\frac{\eta}{2}
\;,
\qquad
\beta\;=\;
2-\eta
\;,
$$
and
$$
A
\;=\;
{\rm sgn}(\epsilon)\;
\frac{\sqrt{|\epsilon\, x|}}{L\,\pi}
\;,
\qquad
B
\;=\;
\frac{1}{8\,L}\,\frac{\EE_\sigma(|x_\sigma|^2)}{|\epsilon\,x|}
\;.
$$

\vspace{.1cm}

\noindent {\rm (ii) (Parabolic regime)} Let $\eta=\frac{4}{3}$, then
$$
\alpha\;=\;\beta\;=\;\frac{2}{3}\;,
$$
and $A=A(\epsilon)$ and $B=B(\epsilon)$ are given by integrals written
out explicitly in {\rm Section~\ref{sec-bandedge}}. 
For $\eta>\frac{4}{3}$, $A$ and
$B$ are independent of $\epsilon$, namely the result is the same as for
$\epsilon=0$. 

\vspace{.2cm}

\noindent {\rm (iii) (Hyperbolic regime)} Let 
$\frac{4}{5}<\eta<\frac{4}{3}$ and let
the sign of $\epsilon\neq 0$ be
such that $E_b+\epsilon\,\lambda^\eta$ is outside the band of
$H_0$. Then
$$
\alpha
\;>\;
\frac{\eta}{2}
\;,
\qquad
\beta\;=\;
\frac{\eta}{2}
\;,
$$
and
$$
B
\;=\;
{\sqrt{|\epsilon\, x|}}
\;.
$$

\end{theo}
\begin{figure}[ht]
\begin{center}
\epsfig{file=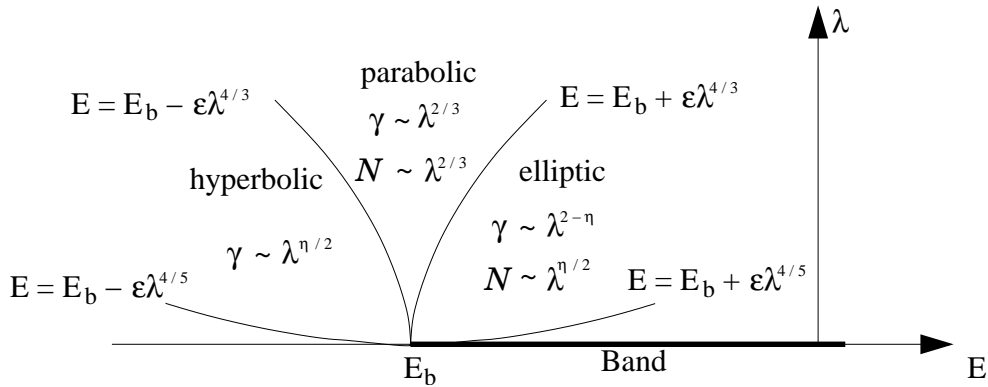, width=13cm} 
\caption{Phase diagram}
\label{fig-phasediagram}
\end{center}
\end{figure}

The result is summarized in Figure~\ref{fig-phasediagram}. All the error
estimates, in particular, the value of $\delta$, are controlled more
explicitely and given in Section~\ref{sec-bandedge}. As will become more 
clear below, the terms elliptic, parabolic and hyperbolic regimes reflect the
nature of the corresponding averaged
transfer matrix to lowest order in $\lambda$.  Let
us point out that we do not provide the asymptotics for the IDS in the
hyperbolic regime, but only give 
an upper bound on it. Actually the IDS between the
Lifshitz tails and the band edge of the periodic operator is exponentially
small in $\lambda$, a fact that results from a more delicate large deviation
behavior. This will be dealt with elsewhere. The Lifshitz tails themselves
were analyzed in \cite{S2}.

\vspace{.2cm}

The proof of Theorem~\ref{theo-scaling} begins (in Section~\ref{sec-normal})
with an adequate basis change on the transfer matrices. A band edge is
characterized by the fact that the modulus of the trace of the 
transfer matrix is $2$. If there is no band touching, the transfer
matrix is then not diagonalizable and its normal form is hence a Jordan
block (Appendix A). The action of a Jordan block on projective space
(in fact, identified with the Pr\"ufer phases)
has exactly one fixed point, which is unstable under random pertubations
induced by the randomness in the Hamiltonian. In order to analyse the
associated invariant distribution on projective space, a further basis
change blowing up the vicinity of the fixed point is necessary. If
this is done adequately, a formal perturbative calculation typically
leads to a differential equation for the invariant distribution. 

\vspace{.2cm}

Instead of studying the concrete form of the random matrices
obtained in Section~\ref{sec-normal} after the various basis changes, 
we rather choose to single out the more general concept of an anomaly
of a family of random matrices ({\it cf.}
Section~\ref{sec-anomalies}). The term anomaly is chosen by reference
to the center of band
anomaly in the one-dimensional Anderson model as studied by 
Kappus and Wegner \cite{KW}. The latter 
is a special case which is analyzed in a prior work \cite{S}. 
The formalism to calculate the Lyapunov exponents and rotation numbers
at a general anomaly is built up in Section~\ref{sec-dynamics}. In
particular, it turns out that one needs to evaluate
certain Birkhoff sums for the perturbative calculation of these quantities. 
The perturbative evaluation of these Birkhoff
sums directly leads to a differential operator on projective space
(Propositions \ref{prop-ellipt} and \ref{prop-DGL}). 
In some situations this operator is of first order, and consequently
the anomaly is called first order as well (this is dealt with in
Section~\ref{sec-elliptic}). However, in the more interesting cases
treated in Sections~\ref{sec-FP} and \ref{sec-2nd},
the differential operator is of second order and of the Fokker-Planck
type. The latter operator was already used in \cite{S} in order 
to study the center of
band anomaly, but here one is
confronted with the supplementary difficulty that its ellipticity
is destroyed at a
band edge. Thus a thorough analysis of the corresponding singularities
is needed. Section~\ref{sec-FP} deals with the Fokker-Planck
operator and its groundstate, while Section~\ref{sec-2nd} shows how it
is used to calculate the Birkhoff sums. Both sections heavily depend
on Appendix B, where inhomogeneous singular first order ordinary
differential equations are studied in detail. 
Sections~\ref{sec-anomalies} up to \ref{sec-2nd} are 
kept slightly more general than needed for the
proof of Theorem~\ref{theo-scaling} as completed in 
Section~\ref{sec-bandedge}, but we hope that this 
stresses the structural aspects
of the analysis. There may possibly also be further applications. 

\vspace{.2cm}

\noindent {\bf Acknowledgment:} We are thankful to Andreas Knauf for numerous
comments and discussions. This work was support by the DFG.

\section{Normal form of transfer matrix at a band edge}
\label{sec-normal}

In this section we motivate and carry out
the basis changes on the transfer matrices in the vicinity of a band
edge. The final normal forms obtained in the various regimes will then
motivate the definition of the term anomaly in the next section.

\vspace{.2cm}

First, one rewrites the Schr\"odinger equation
$H_{\lambda,\omega}\psi=E\psi$, with $E\in\RR$, $\psi\in\ell^2(\ZZ)$
and $H_{\lambda,\omega}$ as given in \eqref{eq-Jacobi}, in the standard
way using transfer matrices ({\it e.g.} \cite{JSS}).
Due to the periodicity of $H_0$
it is convenient to introduce the 
transfer matrix over $L$ sites associated to one configuration
$\sigma\in\Sigma$ of the disorder on these sites:
\begin{equation}
\label{eq-transfer}
\Tt^E_{\lambda,\sigma}
\;=\;
\prod_{l=1}^L
\,
\Tt^E({\hat{t}_l+\lambda\,\tilde{t}_l(\sigma),
\hat{v}_l+\lambda\,\tilde{v}_l(\sigma)})
\;,
\qquad
\Tt^E(t,v)
\;=\;
\left(\begin{array}{cc} (E-v)t^{-1} & -t \\ t^{-1} & 0 
\end{array} \right)
\;.
\end{equation}
Note that $\Tt^E(t,v)\in{\rm SL}(2,\RR)$ and hence 
$\Tt^E_{\lambda,\sigma}\in{\rm SL}(2,\RR)$.
If $E_b$ is a band edge of $H_0$, then $\Tr(\Tt^{E_b}_{0,\sigma})=\pm 2$
and, because there is no band touching ,
$\partial_E\Tr(\Tt^{E_b}_{0,\sigma})\neq 0$. One can show that necessarily 
$\Tt^{E_b}_{0,\sigma}$ has only one eigenvector (see Appendix A). 
Hence one can find a basis change
$N\in{\rm SL}(2,\RR)$ such that $N\Tt^{E_b}_{0,\sigma}N^{-1}$ is a Jordan
block with eigenvalue either $1$ or $-1$. This elementary fact was already
efficiently used for the analysis of the Lifshitz tails \cite{S2}.
Using this basis change, one
can write the full energy rescaled transfer matrix as follows:
\begin{equation}
\label{eq-Qdef}
N\,\Tt^{E_b+\epsilon\lambda^\eta}_{\lambda,\sigma}\,N^{-1}
\;=\;
\pm\;
\exp\left(
\,
\left(\begin{matrix} 
0 & \pm 1 \\ 0 & 0 
\end{matrix} \right)
\;+\; \sum_{k\geq 1}
\lambda^{\eta_k}\,Q_{\eta_k,\sigma}
\;\right)
\;.
\end{equation}
Here the exponents $\eta_k$ are in $\NN+\eta\,\NN$, are 
put into increasing order
$\eta_k<\eta_{k+1}$ and the $Q_{\eta_k,\sigma}$ are in the Lie algebra
${\rm sl}(2,\RR)$. 
Hence the lowest terms are $\lambda^\eta Q_{\eta,\sigma}$ or 
$\lambda Q_{1,\sigma}$, pending on the value of $\eta$. Let us note
that $Q_{k\eta,\sigma}$ are independent of $\sigma$ unless
$k\eta\in\NN$. Also, $\EE_{\sigma}(Q_{1,\sigma})=0$ unless
there is a $k$ such that $k\eta=1$. Each combination of the 
signs can occur, the one inside the exponential indicates
if it is a lower (+ sign) or an upper (- sign) bandedge.
For sake of concreteness, we choose both signs positive. 

\vspace{.2cm}

A few further remarks on the $Q_{\eta_k,\sigma}$ will be relevant
later on. Because there is no band touching at $E_b$, 
for $\epsilon \neq 0$ one gets
\begin{equation}
\label{eq-Qnonvanish}
0
\;\neq\;
\left.\epsilon\,
\partial_E\,\Tr(N\Tt^E_{0,\sigma}N^{-1})
\right|_{E=E_b}
\;=\;
\Tr\left(\,
Q_{\eta,\sigma}
\,
\left(
\begin{matrix} 
1 & 1 \\
0 & 1
\end{matrix}
\right)
\,\right)
\;=\;
\epsilon\,x
\;,
\end{equation}
where the first identity follows from the definition \eqref{eq-Qdef},  
Duhamel's formula and the cyclicity of the trace, and the second one
defines $x$. By the same
calculation (with $\epsilon=0$)
$$
\partial_\lambda\,
\Tr(N\Tt^{E_b}_{\lambda,\sigma}N^{-1})
\,|_{\lambda=0}
\;=\;
\Tr\left(\,
Q_{1,\sigma}
\,
\left(
\begin{matrix} 
1 & 1 \\
0 & 1
\end{matrix}
\right)
\,\right)
\;=\;
x_\sigma
\;,
$$
the latter by definition of the centered random variable $x_\sigma$.
Of course this expression is centered because the perturbation is
centered. However, the assumption in Theorem~\ref{theo-scaling} 
on the non-invariance of the band edge under perturbation translates into
\begin{equation}
\label{eq-QVar}
\EE
\left(\,\left[\Tr\left(\,
Q_{1,\sigma}
\,
\left(
\begin{matrix} 
1 & 1 \\
0 & 1
\end{matrix}
\right)
\,\right)\right]^2
\right)
\;>\;0
\:.
\end{equation}

\vspace{.2cm} 

It turns out (see Section~\ref{sec-dynamics} or \cite{JSS}) that 
one needs to study the random dynamical system given by random iteration of
the natural action of
the transfer matrices $N\Tt^{E_b+\epsilon\lambda^\eta}_{\lambda,\sigma}
N^{-1}$ on  
$S^1_\pi\cong \RR/\pi\ZZ$, given by (see Section~\ref{sec-dynamics}
for details)
\begin{equation}
\label{eq-action0}
e_{\theta_n}
\;=\;\pm\;
\frac{N\,\Tt^{E_b+\epsilon\lambda^\eta}_{\lambda,\sigma_n}\,
N^{-1}e_{\theta_{n-1}}}{
\|N\,\Tt^{E_b+\epsilon\lambda^\eta}_{\lambda,\sigma_n}\,
N^{-1}e_{\theta_{n-1}}\|}
\;,
\qquad
e_\theta
\;=\;
\left(\begin{matrix} 
\cos(\theta) \\ \sin(\theta) 
\end{matrix}\right)
\;.
\end{equation}
%
\begin{figure}[ht]
\begin{center}
\epsfig{file=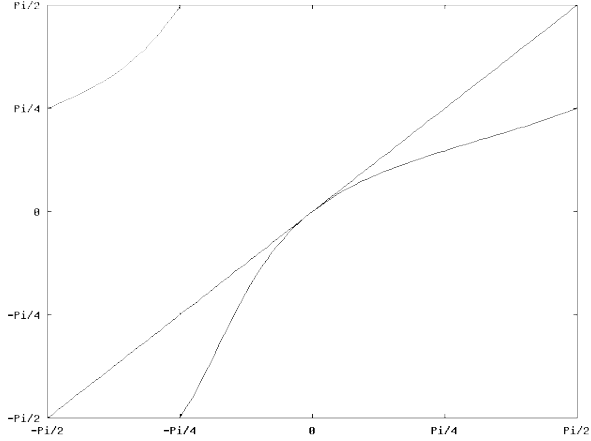,width=8cm} 
\caption{Dynamics at $\lambda=0$ at lower band edge, i.e. + sign in 
{\rm \eqref{eq-Qdef}}.}
\label{fig-freedynamics}
\end{center}
\end{figure}
At $\lambda=0$, the action is given by the graph of Figure~\ref{fig-freedynamics}. It
has an unstable parabolic fixed point at $\theta=0$, which is being approached
from the left under the dynamics without passing it though. 
For $\lambda>0$, the dynamics is the
same to lowest order $\lambda^0$, but the random perturbations of
higher order may allow the dynamics to cross through
$\theta=0$. Nevertheless, the angles $\theta_n$ are close to
$\theta=0$ for most $n$ and thus their distribution is more and more
concentrated in
the vicinity of $\theta=0$ as $\lambda$ tends to $0$ and, in fact,
converges weakly to a Dirac peak. 
In order to extract the shape of the 
non-trivial distribution it is necessary to rescale the neighborhood
of $\theta=0$ with an adequate power of $\lambda$. This is done by
conjugation with
$$
N_{\lambda,\delta}
\;=\;
\left(
\begin{matrix} 
\lambda^\delta & 0 \\
0 & 1
\end{matrix}
\right)
\;.
$$
The conjugation has the following effect on $2\times 2$ matrices:
\begin{equation}
\label{eq-rescale}
N_{\lambda,\delta}
\;
\left(
\begin{matrix} 
a & b \\
c & d
\end{matrix}
\right)
\;
N_{\lambda,\delta}^{-1}
\;=\;
\lambda^{-\delta}\,
\left(
\begin{matrix} 
0 & 0 \\
c & 0
\end{matrix}
\right)
\;+\;
\left(
\begin{matrix} 
a & 0 \\
0 & d
\end{matrix}
\right)
\;+\;
\lambda^{\delta}\,
\left(
\begin{matrix} 
0 & b \\
0 & 0
\end{matrix}
\right)
\:. 
\end{equation}
One deduces from \eqref{eq-Qdef}
$$
N_{\lambda,\delta}
N\,\Tt^{E_b+\epsilon\lambda^\eta}_{\lambda,\sigma}\,N^{-1}
N_{\lambda,\delta}^{-1}
\;=\;
\exp\left(
\,
N_{\lambda,\delta}\left(\begin{matrix} 
0 & 1 \\ 0 & 0 
\end{matrix} \right)N_{\lambda,\delta}^{-1}
\;+\; \sum_{k\geq 1}
\lambda^{\eta_k}\,N_{\lambda,\delta}Q_{\eta_k,\sigma}N_{\lambda,\delta}^{-1}
\;\right)
\;.
$$
Note that the powers of $\lambda$ appearing in the exponential now lie in
$\NN+\eta\,\NN+\{-\delta,0,\delta\}$, but none is negative for
$\delta<\min\{1,\eta\}$. 
The following three contributions of low order in $\lambda$ turn
out to be relevant:
$$
X_{\delta,1}
\;=\;
\lambda^{\delta}\,
\left(
\begin{matrix} 
0 & 1 \\
0 & 0
\end{matrix}
\right)
\,,
\:\:\,
X_{\delta,2}
\;=\;
\lambda^{\eta-\delta}
\lim_{\lambda'\to 0} \lambda'
N_{\lambda',1}\,Q_{\eta,\sigma}\,N_{\lambda',1}^{-1}
\,,
\:\:\,
X_{\delta,\sigma,3}
\;=\;
\lambda^{1-\delta}
\lim_{\lambda'\to 0} \lambda'
N_{\lambda',1}\,Q_{1,\sigma}\,N_{\lambda',1}^{-1}
\,,
$$
where the notation reflects that $Q_{\eta,\sigma}$ is independent of
$\sigma$. 
By equations \eqref{eq-rescale} and \eqref{eq-Qnonvanish}, 
$X_{\delta,2}$ has a non-vanishing constant entry only in the lower
left corner unless $\epsilon=0$. Similarly by 
\eqref{eq-rescale} and \eqref{eq-QVar}, 
$X_{\delta,\sigma,3}$ has a centered random entry in the lower left
corner with positive variance, while all the other entries vanish.
It is helpful to think of $X_{\delta,1}$ and $X_{\delta,2}$ as drift
terms in the action \eqref{eq-action0}, and of $X_{\delta,\sigma,3}$ as
generator of a diffusion.
Now we blow up the parabolic fixed point by augmenting $\delta$ until
the drift $X_{\delta,1}$ is of same order of magnitude as either the
drift $X_{\delta,2}$ (then $\delta=\frac{\eta}{2}$) or the diffusive
force (variance) of $X_{\delta,\sigma,3}$ (then
$2(1-\delta)=\delta$ so that $\delta=\frac{2}{3}$). Hence we choose
$$
\delta
\;=\;
\min\left\{\,\frac{\eta}{2}\,,\,\frac{2}{3}\,\right\}
\;.
$$
Explicitely, for $\eta<\frac{4}{3}$, one has with this choice of $\delta$
\begin{equation}
\label{eq-bandanom}
N_{\lambda,\delta}
N\,\Tt^{E_b+\epsilon\lambda^\eta}_{\lambda,\sigma}\,N^{-1}
N_{\lambda,\delta}^{-1}
\;=\;
\exp
\left(
\,
\lambda^{\frac{\eta}{2}}
\left(
\begin{matrix} 
0 & 1 \\
\epsilon\,x & 0
\end{matrix}
\right)
\,+\,
\lambda^{1-\frac{\eta}{2}}
\left(
\begin{matrix} 
0 & 0 \\
x_\sigma & 0
\end{matrix}
\right)
\,+\,
\lambda^\eta\,Q'_\eta
\,+\,
\lambda\,Q'_{1,\sigma}
\,+\,
\Oo
\,
\right)
\;,
\end{equation}
where $\Oo=\Oo(\lambda^{\frac{3\eta}{2}},\lambda^{1+\frac{\eta}{2}})$ and the
specific form of $Q'_\eta$ and $Q'_{1,\sigma}$ will not be of importance for
the asymptotics of Theorem~\ref{theo-scaling}.
For $\eta>1$, the second term is of lower order than the first, but as
it is centered, we shall see that it does not enter into the
asymptotics of the IDS and the Lyapunov exponent because its variance
is smaller than $\Oo(\lambda^{\frac{\eta}{2}})$. Hence
\eqref{eq-bandanom} is an
anomaly of first order in the sense of the following section. 
In the case $\eta=\frac{4}{3}$,
\begin{equation}
\label{eq-bandanom2}
N_{\lambda,\delta}
N\,\Tt^{E_b+\epsilon\lambda^\eta}_{\lambda,\sigma}\,N^{-1}
N_{\lambda,\delta}^{-1}
\;=\;
\exp
\left(
\,
\lambda^{\frac{1}{3}}
\left(
\begin{matrix} 
0 & 0 \\
x_\sigma & 0
\end{matrix}
\right)
\;+\;
\lambda^{\frac{2}{3}}
\left(
\begin{matrix} 
0 & 1 \\
\epsilon\,x & 0
\end{matrix}
\right)
\;+\;
\Oo(\lambda)
\,
\right)
\;.
\end{equation}
This is actually the same formula as \eqref{eq-bandanom}, but it was written 
out again in order to clearly exhibit the so-called anomaly of second
order (see Definition~\ref{def-critical} below), namely the variance
of the first centered term is precisely of the same order of magnitude as the
expectation value of the second. If $\eta>\frac{4}{3}$, one obtains the same
formula with $\epsilon=0$, but with an additional error term
$\Oo(\lambda^{\eta-\frac{2}{3}})$.

\section{Definition of anomalies}
\label{sec-anomalies}

Let us consider families
$(T_{\lambda,\sigma})_{\lambda\in\RR,\sigma\in\Sigma}$ of
matrices in ${\rm SL}(2,\RR)$ 
depending on a random variable $\sigma$ in some probability space
$(\Sigma,{\bf p})$ as well as a real coupling parameter
$\lambda$. In order to avoid technicalities, 
we suppose that $T_{\lambda,\sigma}$ has compact support for small $\lambda$.
Furthermore we assume that the dependence on $\lambda$ can be expanded 
in some power series with non-negative exponents (not necessarily integers). 
Later on, we shall choose $T_{\lambda,\sigma}$ to be the matrices in
\eqref{eq-bandanom} and \eqref {eq-bandanom2}.

\begin{defini}
The value $\lambda=0$ is an anomaly  of the family 
$(T_{\lambda,\sigma})_{\lambda\in\RR,\sigma\in\Sigma}$ if for all
$\sigma\in\Sigma$ and a sign that may depend on $\sigma\in\Sigma$:
\begin{equation}
\label{eq-sign}
T_{0,\sigma}\;=\;\pm\,{\bf 1}
\mbox{ . }
\end{equation}
\end{defini}

In order to
further classify the anomalies, let us suppose that for a fixed 
$\lambda$- and 
$\sigma$-independent basis change $M\in \mbox{\rm SL}(2,\RR)$
to be chosen later, the transfer matrix is of the following form:
\begin{equation}
\label{eq-expan}
MT_{\lambda,\sigma}M^{-1} 
\;=\;
\pm\,\exp\left(\sum_{k\geq 1}
\lambda^{\eta_k}\,P_{\eta_k,\sigma}\;\right)
\;.
\end{equation}
Here
$P_{\eta_k,\sigma}\,\in\,\mbox{\rm sl}(2,\RR)$,  
$\eta_k>0$ for $k\in \NN$ and $K$, such that 
$\eta_j < \eta_k$ if $j<k$, 
$\eta_{K} = 2\eta_1$, $\eta_{K+1} \leq \eta_1+\eta_2$ and 
$\pp(P_{\eta_k,\sigma}= 0)<1$ for $k=1,\ldots,K-1$ 
{\rm (}which means none of these matrices shall be
identically 0 for $\pp$-almost all $\sigma${\rm )}. 

\begin{defini}
\label{def-critical}
Let $(T_{\lambda,\sigma})_{\lambda\in\RR,\sigma\in\Sigma}$ have an
anomaly and suppose given the expansion {\rm \eqref{eq-expan}} for a
fixed basis change $M$.
The anomaly is said to be of first order and $\Kk$th 
kind if $\Kk< K$, 
$\EE(P_{\eta_k,\sigma})=0$ for $k=1,\ldots,\Kk-1$ and 
$\EE(P_{\eta_\Kk,\sigma})$ 
is non-vanishing.
An anomaly of first order and $\Kk$th kind is called elliptic 
if $\det(\EE(P_{\eta_\Kk,\sigma}))>0$, hyperbolic if 
$\det(\EE(P_{\eta_\Kk,\sigma}))<0$
and parabolic if $\det(\EE(P_{\eta_\Kk,\sigma}))=0$. 
Note that all these notions are independent of the choice of $M$.

\vspace{.2cm}

If $\;\EE(P_{\eta_k,\sigma})=0$, for $k=1,\ldots, K-1$ 
{\rm (}then the variance of $P_{\eta_1,\sigma}$ is 
non-vanishing{\rm )}, 
then an anomaly is said to be of second order.
\end{defini}

It may happen that the transfer matrices have to be modified as
follows in order to obtain an anomaly.
If there is a mapping $\lambda\mapsto M_\lambda\,\in\,{\rm GL}(2,\RR)$ 
for $\lambda>0$ where $M_\lambda$ is independent of $\sigma$, 
such that
\begin{equation}
\label{eq-transform}
\lim_{\lambda\to 0} M_\lambda\,T_{\lambda,\sigma}\,M^{-1}_\lambda
\;=\;
\pm\,{\bf 1} 
\;,
\end{equation}

\noindent then we say that $T_{\lambda,\sigma}$ is transformed to an 
anomaly by  $M_\lambda$.
Note that the limits $\lim_{\lambda\downarrow 0} M_\lambda$ and 
$\lim_{\lambda\downarrow 0} M_\lambda^{-1}$ need not exist.

\vspace{.2cm}

As argued in the last section,
the transfer matrices at a band edge can be transformed into anomalies
\eqref{eq-bandanom} and \eqref{eq-bandanom2} of
respectively first and second order. The term {\it anomaly} first
appeared in the work of Kappus and Wegner \cite{KW} on the band center
of a one-dimensional Anderson model. Indeed, as discussed in \cite{S}, 
the square of the
transfer matrix at a band edge is an anomaly in the above sense (of
second order, if the random potential is centered). More generally,
higher powers of transfer matrices are transformed into anomalies for
rational quasimomenta, and thus lead to anomalies in
higher order perturbation theory \cite{DG}. A more systematic
classification of anomalies as in Definition~\ref{def-critical} was
done in \cite{S}. However, we felt it adequate to change the term {\it
degree of an anomaly} used in \cite{S} to the present
{\it order of an anomaly} because it leads to a
differential equation of corresponding order.

\section{Phase shift dynamics}
\label{sec-dynamics}

The bijective action $\Ss_T$ of a matrix $T\in\mbox{SL}(2,\RR)$ on 
$S^1_\pi= \RR/\pi\ZZ=[0,\pi)$ is given by

\begin{equation}
\label{eq-action}
e_{\Ss_T(\theta)}
\;=\;
\pm\;\frac{Te_{\theta}}{
\|Te_{\theta}\|}
\;,
\qquad
\;\;\;\;\;
e_\theta
\;=\;
\left(
 \begin{array}{cc} \cos(\theta) \\ \sin(\theta)
\end{array}
\right)
\mbox{ , }
\;\;
\theta\in S^1_\pi
\mbox{ , }
\end{equation}

\noindent with an adequate choice of the sign. 
This defines a group action, namely
$\Ss_{TT'}=\Ss_{T}\Ss_{T'}$. In particular the map $\Ss_T$ is invertible
and $\Ss_T^{-1}=\Ss_{T^{-1}}$. Note that this is an action on
projective space. 
In order to shorten notations, we write 
$$
\Ss_{\lambda,\sigma}
\;=\;
\Ss_{MT_{\lambda,\sigma}M^{-1}}
\;.
$$
At an anomaly one has
$\Ss_{\lambda,\sigma}(\theta)=\theta+\Oo(\lambda)$. 

\vspace{.2cm}

Given an initial angle $\theta_0$ and iterating this dynamics defines
a Markov process $\theta_n(\omega,\theta_0)$ on
$\Omega=\Sigma^\ZZ\times\{1,\ldots,L\}$, {\it i.e.} 
for $\omega=((\sigma_n)_{n\in\ZZ},k)$ one defines iteratively
\begin{equation}
\label{eq-randomdyn}
\theta_0(\omega,\theta_0)
\;=\;
\theta_0
\,,
\qquad
\theta_{n+1}({\omega,\theta_0})
\;=\;
S_{{\lambda,\sigma_{n+1}}}(\theta_n(\omega,\theta_0))
\;.
\end{equation}
In order to shorten notations, we denote 
$\theta_n(\omega,\theta_0)$ by $\theta_n$.
Under adequate identifications, this corresponds to the dynamics
\eqref{eq-action0}. The $\theta_n$ are also called modified Pr\"ufer
variables. 

\vspace{.2cm}

In order to analyze the dynamics in more detail, some further
notations are needed. For $k\in\NN$, we define the 
trigonometric polynomials:
$$
p_{k,\sigma}(\theta)
\;=\;
\Im m \left(\frac{\langle v|P_{\eta_k,\sigma}|e_\theta\rangle}{\langle
v|e_\theta\rangle}
\right)
\;,
\qquad 
v\;=\;
\frac{1}{\sqrt{2}}\left(\begin{array}{c}
1\\-\imath\end{array}\right)
\;.
$$
More explicitly, in terms of the matrix elements one obtains
\begin{equation}
\label{eq-matrixnotation}
P_{\eta_k,\sigma}
\;=\;
\left(\begin{matrix} a_{\eta_k,\sigma} & b_{\eta_k,\sigma} \\
c_{\eta_k,\sigma} & -a_{\eta_k,\sigma}
\end{matrix}\right)
\quad
\Longrightarrow
\quad
p_{k,\sigma}(\theta)
\;=\;
c_{\eta_k,\sigma}\cos^2(\theta)\,-\,b_{\eta_k,\sigma}\sin^2(\theta)
\,-\,a_{\eta_k,\sigma}\sin(2\theta)
\;.
\end{equation}

Now starting from the identity 

$$
e^{2\imath \Ss_{\lambda,\sigma}(\theta)}
\;=\;
\frac{\langle v|MT_{\lambda,\sigma} M^{-1}|e_\theta\rangle}{\langle 
\overline{v}|MT_{\lambda,\sigma}
M^{-1}|e_\theta\rangle}
\;,
$$

\noindent the definition (\ref{eq-expan}) and the identity
$\langle v|e_\theta\rangle=\frac{1}{\sqrt{2}}\;e^{\imath\theta}$, one can
verify that 

$$
\Ss_{\lambda,\sigma}(\theta)
\;=\;
\theta\,+\,
\Im m 
\left(\sum_{k=1}^{K}
\lambda^{\eta_k}\,\frac{\langle v|P_{\eta_k,\sigma}|e_\theta\rangle}{\langle
v|e_\theta\rangle}
\,+\,
\lambda^{2\eta_1}
\left[ \frac{\langle v|P_{\eta_1,\sigma}^2|e_\theta\rangle}{2\,\langle
v|e_\theta\rangle}
\,-\,
\,\frac{\langle v|P_{\eta_1,\sigma}|e_\theta\rangle^2}{2\,\langle
v|e_\theta\rangle^2}\right]
\right)
\,+\,\Oo(\lambda^{\eta_{K+1}})
\;.
$$
\noindent As one readily verifies that 
$$
P^2_{\eta_1,\sigma}
\;=\;
-\det(P_{\eta_1,\sigma})\; {\bf 1}\;,
\qquad
\Im m 
\left(
\frac{\langle v|P_{\eta_1,\sigma}|e_\theta\rangle^2}{\langle
v|e_\theta\rangle^2}
\right)
\;=\;
-\,p_{1,\sigma}\,
\partial_\theta\, p_{1,\sigma}(\theta)
\;,
$$
\noindent it follows that
\begin{equation}
\label{eq-polydef}
\Ss_{\lambda,\sigma}(\theta)
\;=\;
\theta\,+\,\sum_{k=1}^K \lambda^{\eta_k}\,p_{k,\sigma}(\theta)
\,+\,\frac{1}{2}\,\lambda^{2\eta_1}\,
p_{1,\sigma}\,
\partial_\theta\, p_{1,\sigma}(\theta)
\,+\,\Oo(\lambda^{\eta_{K+1}})
\;.
\end{equation}
\noindent As one has
$\exp\left(\sum \lambda^{\eta_k}\,P_{\eta_k,\sigma}
\right)^{-1}=
\exp\left(-\sum \lambda^{\eta_k}\,P_{\eta_k,\sigma}
\right)$ the same procedure leads to
\begin{equation}
\label{eq-polyinv}
\Ss_{\lambda,\sigma}^{-1}(\theta)
\;=\;
\theta\,-\,\sum_{k=1}^K \lambda^{\eta_k}\,p_{k,\sigma}(\theta)\,+\,
\frac{1}{2}\,\lambda^{2\eta_1}\,
p_{\eta_1,\sigma}\,
\partial_\theta \,p_{\eta_1,\sigma}(\theta)
\,+\,\Oo(\lambda^{\eta_{K+1}})
\;.
\end{equation}

Our main interest is the perturbative calculation of 
the Lyapunov exponent $\gamma(\lambda)$ and rotation number
$\Rr(\lambda)$ associated to the 
random family of matrices  $(T_{\lambda,\sigma_n})_{n\geq 1}$ which 
are defined by
\begin{equation}
\label{eq-lyap}
\gamma(\lambda) 
\;=\; 
\lim_{N \to\infty}\,\frac{1}{N}
\,\EE\,\sum_{n=0}^{N-1}
\EE_\sigma\,\log(\|MT_{\lambda,\sigma}M^{-1} e_{\theta_n}\|) \;,
\end{equation}
and
\begin{equation}
\label{eq-rot}
\Rr(\lambda)
\;=\;
\frac{1}{\pi}\,
\lim_{N\rightarrow\infty}\,
\frac{1}{N}\,\EE\,\sum_{n=1}^N 
\EE_\sigma \,\varphi_{\lambda,\sigma}(\theta_{n})
\;.
\end{equation}
where $\varphi_{\lambda,\sigma}:S_\pi^1\to\RR$ is the phase 
shift given by $\varphi_{\lambda,\sigma}(\theta)=
\Ss_{\lambda,\sigma}(\theta)-\theta$. If $T_{\lambda,\sigma}=
\tilde{N}\Tt_{\lambda,\sigma}^{E+\epsilon\lambda^\eta}\tilde{N}^{-1}$ 
where the latter is the
transfer matrix as given in \eqref{eq-transfer} associated to a random Jacobi
matrix $H_{\lambda,\omega}$ with an $L$-periodic background operator defined in
\eqref{eq-Jacobi} and $\tilde{N}$ is an arbitrary basis change such 
as the ones used in \eqref{eq-bandanom} and
\eqref{eq-bandanom2}, then the Lyapunov exponent $\gamma_\lambda(E)$ (inverse
localization length) and IDS $\Nn_\lambda(E)$ of the random Jacobi matrix are 
\begin{equation}
\label{eq-links}
\gamma_\lambda(E+\epsilon\lambda^\eta)
\;=\;
\frac{1}{L}\;\gamma(\lambda)
\;,
\qquad
\Nn_\lambda(E+\epsilon\lambda^\eta)
\;=\;
-\;\frac{1}{L}
\,
\Rr(\lambda)
\!\!\!\mod\frac{1}{L}
\;.
\end{equation}
The first identity follows immediately from the definition; the second is a
consequence of the oscillation theorem and the gap labelling for the periodic
operator (see {\it e.g.} \cite{JSS} for
details). In both of the expressions \eqref{eq-lyap} and 
\eqref{eq-rot}, one can now expand
each summand w.r.t. $\lambda$.

\begin{lemma}
\label{lem-coefficients}
Set
$$
\alpha_{\eta,\sigma}
\;=\;
\langle v|\,P_{\eta,\sigma}\,|v\rangle
\;,
\qquad
\beta_{\eta,\sigma}
\;=\;
\langle \overline{v}|\,P_{\eta,\sigma}\,|v\rangle
\;,
\qquad
\gamma_{\eta,\sigma}
\;=\;
\langle \overline{v}|\,|P_{\eta,\sigma}|^2\,|v\rangle
\;
$$
as well as
$$
\delta_{\eta_1,\eta_2,\sigma}
\;=\;
\frac{1}{2}\,
\langle \overline{v}|\,(P_{\eta_1,\sigma}\,+\,P_{\eta_1,\sigma}^*) 
P_{\eta_2,\sigma}
\,+\,(P_{\eta_2,\sigma}\,+\,P_{\eta_2,\sigma}^*) P_{\eta_1,\sigma}\,|
v \rangle\;.
$$
Then $p_{\eta,\sigma}(\theta)=\Im m(
\alpha_{\eta,\sigma}-\beta_{\eta,\sigma}e^{2\imath\theta})$. Furthermore,
with errors of order $\Oo(\lambda^{3\eta_1})$,
\begin{eqnarray}
\log(\|MT_{\lambda,\sigma}M^{-1} e_{\theta}\|)
& = &
\Re e\,\sum_{k\geq 1}
\Bigl(
\lambda^{\eta_k} \,
\beta_{\eta_k,\sigma}\,e^{2\imath\theta}
\,+\,
\frac{\lambda^{2\eta_k}}{2}\,
\bigl(
|\beta_{\eta_k,\sigma}|^2
\,+\,
\gamma_{\eta_k,\sigma}\,e^{2\imath\theta}
\,-\,
\beta_{\eta_k,\sigma}^2\,e^{4\imath\theta}
\bigr)
\Bigr)
\,+
\nonumber
\\
& &
\label{eq-expan1}
\\
& + & 
\Re e\,\sum_{k_1 < k_2}
\Bigl(
\lambda^{\eta_{k_1}+\eta_{k_2}} \,
\bigl(
\beta_{\eta_{k_1},\sigma}\overline{\beta_{\eta_{k_2},\sigma}}
\,+\,
\delta_{\eta_{k_1},\eta_{k_2},\sigma}\,e^{2\imath\theta}
\,-\,
\beta_{\eta_{k_1},\sigma}\,\beta_{\eta_{k_2},\sigma}\,e^{4\imath\theta}
\bigr)
\Bigr),
\nonumber
\end{eqnarray}
and up to errors of order $\Oo(\lambda^{\eta_1+\eta_2})$ 
\begin{equation}
\label{eq-expan2}
\varphi_{\lambda,\sigma}(\theta)
\;=\;
\Im m
\,\sum_{k\geq 1}\Bigl(
\lambda^{\eta_k} \,
\left(
\alpha_{\eta_k,\sigma}\,-\,
\beta_{\eta_k,\sigma}\,e^{2\imath\theta}
\right)
\,+\,
\frac{1}{2}\,
\lambda^{2\eta_k}\,
\left(
-2\,\alpha_{\eta_k,\sigma}\,\beta_{\eta_k,\sigma}\,
e^{2\imath\theta}
\,+ \,
\beta_{\eta_k,\sigma}^2\,e^{4\imath\theta}
\right)
\Bigr)
\;.
\end{equation}
\end{lemma}

\noindent {\bf Proof.} This follows from straightforward algebra using 
$$ 
\langle e_\theta | T | e_\theta \rangle 
\;=\; 
\frac{1}{2}\, \Tr(T)\,+\,\Re e \left( \
\langle \bar{v} | T | v \rangle \;e^{2\imath\theta} \right)\;,
\qquad 
\Tr(|P_{\eta,\sigma}|^2 + P_{\eta,\sigma}^2) 
\;=\; 4 |\beta_{\eta,\sigma}|^2
\;,
$$
as well as the
identities $\Tr(P_{\eta,\sigma})=0$ and 
$ \langle\bar{v}|P_{\eta,\sigma}^2|v\rangle=0$.
\hfill $\Box$

\vspace{.2cm}

Once these formulas are replaced in \eqref{eq-lyap} and 
\eqref{eq-rot}, one hence needs to consider
Birkhoff sums of the type
\begin{equation}
\label{eq-osci}
I_N(f)
\;=\;
\frac{1}{N}\,\EE\,\sum_{n=0}^{N-1} f(\theta_n)
\;,
\qquad
I(f)\;=\;
\lim_{N\to\infty}\,I_N(f)
\;,
\end{equation}
\noindent for $\pi$-periodic functions $f$. For $\gamma(\lambda)$ and
$\Rr(\lambda)$ one actually only needs the functions
$e^{2\imath\theta}$ and $e^{4\imath\theta}$. These Birkhoff sums have to be
evaluated 
perturbatively in $\lambda$ with a rigorous control on the error
terms. For this purpose, one needs to know the distribution of
the $\theta_n$ as generated by the random dynamics
\eqref{eq-polydef}. This is the main focus of the next sections.

\section{First order anomalies}
\label{sec-elliptic}

It turns out that it is easiest to calculate the Birkhoff sums
$I_N(f)$ in case of an
elliptic first order anomaly of $\Kk$th kind. 
As then $\det(\EE(P_{\eta_\Kk,\sigma}))>0$, one can choose the basis change
$M$ in \eqref{eq-expan} such that 
\begin{equation}
\label{eq-ellipchange}
\EE(P_{\eta_\Kk,\sigma})
\;=\;
\left(\begin{matrix}  
0 & -\,\mu
\\
\mu & 0
\end{matrix}\right)
\;,
\end{equation}
with $\mu\neq 0$. Hence $\EE(p_{\Kk,\sigma}(\theta))=\mu$.

\begin{proposi}
\label{prop-ellipt}
Let $T_{\lambda,\sigma}$ have an elliptic anomaly of first order
and $\Kk$th kind and suppose that $\EE(p_{\Kk,\sigma}(\theta))=\mu$. 
Then for any $f\in C^1(S^1_\pi)$, one has 
$$
I_N(f)
\;=\;
\int_0^\pi \frac{{\rm d}\theta}{\pi}\;f(\theta)
\,+\, 
\Oo\left(\lambda^{\eta_{\Kk'}-\eta_\Kk},
(\lambda^{\eta_\Kk}N)^{-1} \right)\;,
$$
with $\Kk'=\min\{\tilde{\Kk},K\}>\Kk$ where
$\tilde{\Kk}$ is such that $P_{\eta_{\tilde{\Kk}}}$ is the first 
uncentered term in {\rm \eqref{eq-expan}} after $P_{\eta_{\Kk}}$.
\end{proposi}

\noindent {\bf Proof.}
Because $I_N(f)=c+I_N(f-c)$
for $c=\int_0^\pi {\rm d}\theta\;f(\theta)/\pi$, we may assume that 
$\int_0^\pi {\rm d}\theta\;f(\theta)=0$. Then $f$ has an
antiderivative $F\in C^2(S^1_\pi)$. Using the Taylor expansion and
$2\,\eta_1=\eta_K$,
$$ 
F(\Ss_{\lambda,\sigma}(\theta))
\, = \,
F\left(\theta+\sum_{k=1}^K \lambda^{\eta_k} \,p_{k,\sigma}(\theta)
\,+\,
\Oo(\lambda^{\eta_{K+1}})\right) 
\, =\,
F(\theta)\,+\,F'(\theta)
\sum_{k=1}^{K-1} 
\lambda^{\eta_k}\,p_{k,\sigma}(\theta)
\,+\,\Oo(\lambda^{\eta_{K}})
\,.
$$
From this one deduces after taking expectation and summing over $n$:
$$
I_N(F)
\;=\;
I_N(F)\;+\;\mu\,\lambda^{\eta_{\Kk}}\;I_N(f)\;
+\;
\Oo(\lambda^{\eta_{\Kk'}},N^{-1})
\;.
$$
As $\mu\neq 0$ this proves the estimate.
\hfill $\Box$ 

\vspace{.2cm}

Concerning hyperbolic and parabolic first order anomalies, we do not provide
an exhaustive treatment, but rather present some procedures on how these
anomalies can (possibly) be transformed into a more accessible anomaly such as
one of second order. This applies, in particular, to the hyperbolic first order
anomaly \eqref{eq-bandanom} with $\epsilon x$ negative, corresponding to the
hyperbolic regime in Theorem~\ref{theo-scaling}. For a hyperbolic 
first order anomaly of $\Kk$th kind, let us first choose $M$ such that
\begin{equation}
\label{eq-hypchange}
\EE(P_{\eta_\Kk,\sigma})
\;=\;
\left(\begin{matrix} 
-\mu & 0 \\ 0 & \mu
\end{matrix}\right)
\;,
\qquad
\mu>0\;.
\end{equation}
Then to lowest order in $\lambda$, 
$\theta=0$ is the unstable and $\theta=\frac{\pi}{2}$
the stable fixed point of (the averaged dynamics of) $\Ss_{\lambda,\sigma}$.
Like in \eqref{eq-bandanom2}, we use the 
$\lambda$-dependent basis change $N_{\lambda,\delta}$.
Let $\chi$ be the smallest exponent such that $\EE(c_{\chi,\sigma}^2) > 0$.
First suppose that $\EE(c_{\xi,\sigma})=0$ for all 
$\xi<\frac{1}{2}\eta_\Kk+\chi$, then choose
$\delta=\chi-\frac12\eta_\Kk>0$ so that $ 2(\chi-\delta) = \eta_\Kk$. One
obtains 
\begin{equation}
\label{eq-hyptransform}
N_{\lambda,\delta}\,MT_{\lambda,\sigma}M^{-1}\,N_{\lambda,\delta}^{-1} 
\;=\;
\pm \exp\left[\lambda^{\frac{1}{2}\eta_\Kk}\left(\begin{matrix} 
0 & 0 \\ c_{\chi,\sigma} & 0 \end{matrix}\right) 
+ \ldots + 
\lambda^{\eta_\Kk} \left(\begin{matrix} 
a_{\eta_\Kk,\sigma} & b_{\eta_\Kk-\delta,\sigma} \\ 
c_{\eta_\Kk+\delta,\sigma} & -a_{\eta_\Kk,\sigma} 
\end{matrix}\right) + \ldots \right]\;,
\end{equation}
where, moreover, $\EE(b_{\eta_\Kk-\delta,\sigma})=0$ and
$\EE(a_{\eta_\Kk,\sigma})=-\mu<0$. This is a second order anomaly
for which the Birkhoff sums are analyzed in Theorem \ref{theo-Birkhoff2}
below.  If, on the other hand,
there is a $\xi<\frac{1}{2}\eta_\Kk+\chi$ such 
that $\EE(c_{\xi,\sigma}) \neq 0$ and $\xi$ is the 
smallest such exponent, choose $\delta=\xi-\eta_\Kk > 0$ so that 
$$
N_{\lambda,\delta}\,MTM^{-1}\,N_{\lambda,\delta}^{-1} 
\;=\;
\pm \exp\left[ \lambda^{\zeta} 
\left(\begin{matrix} 
a_{\zeta,\sigma} & 0 \\ c_{\zeta+\delta,\sigma} & -a_{\zeta,\sigma} 
\end{matrix}\right) + \ldots + 
\lambda^{\eta_\Kk} \left(\begin{matrix} 
a_{\eta_\Kk,\sigma} & b_{\eta_\Kk-\delta,\sigma} 
\\ c_{\xi,\sigma} & -a_{\eta_\Kk,\sigma} 
\end{matrix}\right) \right]\;
$$
where $\zeta$ is either $\eta_1$ or 
$\chi-\delta$. For both cases we have $2\,\zeta>\eta_\Kk$. 
As $ \EE(b_{\eta_\Kk-\delta,\sigma})=0 $, one thus has again
a hyperbolic first degree anomaly. Repeating the procedure, 
one may now be in the above advantageous case.  

\vspace{.2cm}

A parabolic first order anomaly of $\Kk$th kind can be treated 
just as a the band edge (which is parabolic to $0$th order) in
Section~\ref{sec-normal}. One first chooses
$M$ to be a rotation matrix, such that the parabolic fixed point 
is $\theta=0$, namely $\EE(P_{\eta_\Kk,\sigma})e_0=0$, and then 
considers the matrices
$N_{\lambda,\delta}\,MT_{\lambda,\sigma}M^{-1}\,N_{\lambda,\delta}^{-1}$.
Let again $\chi$ be the smallest exponent such 
that $\EE(c_{\chi,\sigma}^2)>0$ and let $\xi$ be the smallest exponent 
such that $\EE(c_{\xi,\sigma})\neq 0$.
If $\xi \geq \frac{4}{3}\chi+\frac{1}{3}\eta_\Kk$ and 
$\EE(a_{\eta,\sigma})=0$ for all $\eta < \frac{2}{3}(\chi+\eta_\Kk)$,
then one chooses $\delta=\frac{1}{3}(2\chi-\eta_\Kk)>0$ in order 
to get a second order anomaly
$$ 
\pm \exp\left[ \lambda^{\frac13(\chi+\eta_\Kk)}
\left(\begin{matrix} 
a_{\chi-\delta,\sigma} & 0 \\ c_{\chi,\sigma} & -a_{\chi-\delta,\sigma} 
\end{matrix}\right) + \ldots + 
\lambda^{\frac23(\chi+\eta_\Kk)}
\left(\begin{matrix} 
a_{\eta_\Kk+\delta,\sigma} & b_{\eta_\Kk,\sigma} 
\\ c_{\eta_\Kk+2\delta,\sigma} &
-a_{\eta_\Kk+\delta,\sigma} 
\end{matrix}\right) 
\;+\;\ldots \right]\;. 
$$ 

\section{Fokker-Planck operator of a second order anomaly}
\label{sec-FP}

In order to calculate $I_N(f)$ at a second order anomaly, we first
refine the proof of Proposition~\ref{prop-ellipt}. This
naturally leads to an associated differential operator, 
as shown in the next proposition. The remainder of this section
analyzes the properties of these operators.

\begin{proposi}
\label{prop-DGL}
Let the random family $(T_{\lambda,\sigma})_{\sigma\in\Sigma}$ have a
second order anomaly and $F\in C^3(S^1_\pi)$. Then, for
\begin{equation}
\label{eq-inhomODE}
f
\;=\;
\EE(p_{1,\sigma}^2)F''\,+\,\bigl(2\,\EE(p_{K,\sigma})+
\EE(p_{1,\sigma}\partial_\theta p_{1,\sigma})\bigr)
F'
\;,
\end{equation}
one has
$$
I_N(f)
\;=\;
\Oo\left(\lambda^{\tilde{\eta}-\eta_K},(\lambda^{\eta_K}N)^{-1}\right)
\;.
$$
with $\tilde\eta=\min\{\eta_{\tilde{K}},\eta_1+\eta_2\}$ where
$\tilde{K}$ is the first uncentered term in {\rm \eqref{eq-expan}}
after $P_{\eta_K,\sigma}$.
\end{proposi}

\noindent {\bf Proof.} A
Taylor expansion implies
\begin{eqnarray*}
& F & \!\!\!\!\!(S_{\lambda,\sigma}(\theta))
\, = \;
F\left(\theta+\sum_{k=1}^{\tilde{K}-1} \lambda^{\eta_k} \,p_{k,\sigma}(\theta)
\,+\,
\frac{1}{2}\,\lambda^{2\eta_1}
\,p_{1,\sigma}\partial_\theta p_{1,\sigma}(\theta)
\,+\,
\Oo(\lambda^{\tilde{\eta}})\right) 
\\
&  &  \!\!\! =\;
F(\theta)\,+\,F'(\theta)
\sum_{k=1}^{\tilde{K}-1} 
\lambda^{\eta_k}\,p_{k,\sigma}(\theta)
\,+\,
\frac{1}{2}\,\lambda^{2\eta_1}
\,
\left(p_{1,\sigma}\partial_\theta p_{1,\sigma}(\theta)
F'(\theta)+p_{1,\sigma}^2(\theta)F''(\theta)
\right)
\,+\, \Oo(\lambda^{\tilde{\eta}})
\,.
\end{eqnarray*}
Taking expectation values and summing over $n$, one gets the estimate
by the same argument as in Proposition~\ref{prop-ellipt} because
$\eta_K=2\,\eta_1$.
\hfill $\Box$ 

\vspace{.2cm}

Let us sketch the further strategy.
In view of \eqref{eq-inhomODE}, 
it is natural to introduce:
\begin{equation}
\label{eq-polynomials}
p
\;=\;
\EE(p_{1,\sigma}^2)
\;,
\qquad
q
\;=\;
2\,\EE(p_{K,\sigma})\;+\;\EE(p_{1,\sigma}
\,\partial_\theta\,p_{1,\sigma})
\;.
\end{equation}
Both $p$ and $q$ are  the trigonometric
polynomials of degree $4$ on $S^1_\pi$, namely linear combinations of
$e^{\pm 4\imath\theta}$, $e^{\pm 2\imath\theta}$ and the constant
function. Moreover, $p\geq 0$. Hence $p$ can have at most one 
zero of order $4$, or two zeros of order $2$. Actually a zero $\hat{\theta}$
of order $4$
only appears in the rather special case described next. 
By applying a rotation matrix
$M$ in \eqref{eq-expan}, one may assume $\hat{\theta}=\frac{\pi}{2}$ and then
one readily checks that
$$
p(\theta)
\;=\;
\EE(c_{\eta_1,\sigma}^2)\,\cos^4(\theta)\;,
\qquad
P_{\eta_1,\sigma}
\;=\;
\left(
\begin{array}{cc}
0 & 0
\\
c_{\eta_1,\sigma} & 0
\end{array}
\right)
\;.
$$
This is precisely the situation in 
\eqref{eq-bandanom2}. 
Furthermore, let us note that $q$ can have
zeros with orders adding up to at most $4$. 

\vspace{.2cm}

Proposition~\ref{prop-DGL} now
states that one can control Birkhoff sums of functions
which are in the range of the differential operator on $C^3(S^1_\pi)$ 
given by
$$
\Ll
\;=\;
(p\,\partial_\theta\,+\,q)\,
\partial_\theta
\;.
$$
The formal adjoint on $L^2(S^1_\pi)$, also considered as an operator
defined on $C^3(S^1_\pi)$, is given by
$$
\Ll^*
\;=\;
\partial_\theta\,
(\partial_\theta\,p\,-\,q)
\;.
$$
The operator $\Ll^*$ is a Fokker-Planck (or forward Kolmogorov) 
operator on $S^1_\pi$ describing a drift-diffusion dynamics 
on $S^1_\pi$, while $\Ll$ is the associated backward Kolmogorov
operator \cite{Ris}. The operator $\Ll^*$ 
was already used in \cite{S}, but only in
the situation of a strictly positive $p$. 
In the case of the anomaly \eqref{eq-bandanom2} corresponding to the parabolic
regime at a band edge, $p$ has a zero. Hence the ellipticity
of $\Ll^*$ is destroyed. This is the main technical problem that has to be
dealt with in the present work.

\vspace{.2cm}

As Ran$(\Ll)\subset\,$Ker$(\Ll^*)^\perp$, a
necessary condition for controlling the Birkhoff sum $I_N(f)$ 
is hence that $f\in\,$Ker$(\Ll^*)^\perp$. 
As will shortly be shown in Theorem~\ref{theo-FP}, in many situations
Ker$(\Ll^*)$ is one-dimensional and spanned by
a smooth positive, $L^1$-normalized 
function $\rho$. Let us call $\rho$ the groundstate of the Fokker-Planck
operator, even though the spectrum of the Fokker-Planck operator is
non-positive. Furthermore, the
necessary condition $f\in\,$Ker$(\Ll_j^*)^\perp$ combined with 
$f\in  C^2(S^1_\pi)$ will in most cases
turn out (Theorem~\ref{theo-Birkhoff} below)
to be sufficient for
finding a solution $F\in C^3(S^1_\pi)$ of the inhomogeneous
differential equation \eqref{eq-inhomODE}. Thus one deduces 
from Proposition~\ref{prop-DGL} that, for $f\in  C^2(S^1_\pi)$,
$$
I_N(f)
\;=\;
\int {\rm d}\theta\;\rho(\theta) \; f(\theta)
\;+\;
\Oo\left(\lambda^{\eta_{K+1}-\eta_K},(\lambda^{\eta_K}
N)^{-1}\right)\,. 
$$
In some other situations, relevant for the hyperbolic regime in
Theorem~\ref{theo-scaling}, the kernel of $\Ll^*$ is
given by a Dirac peak $\rho$
(if $\Ll^*$ is considered as an operator on the
space of distributions), and the same formula holds.

\vspace{.2cm}

Before proving the main result on the groundstate $\rho$, let us 
point out that $\rho$ can be seen as  
the lowest order formal approximation to the
Furstenberg measure $\nu_\lambda$ on $S^1_\pi$ associated to family
$(T_{\lambda,\sigma})_{\sigma\in\Sigma}$ of random matrices. It is
defined by \cite{BL}
\begin{equation}
\label{eq-invariant}
\int_0^\pi \nu_\lambda({\rm d}\theta)\,f(\theta)
\;=\;
\EE\,
\int_0^\pi 
\nu_\lambda({\rm d}\theta)\,f(S_{\lambda,\sigma}(\theta))
\mbox{ , }
\qquad
f\in C(S^1_\pi)
\mbox{ . }
\end{equation}
\noindent In fact, supposing
$\nu_\lambda( {\rm d}\theta)=\rho_\lambda(\theta)\, {\rm d}\theta$, 
\eqref{eq-invariant} leads to
$$ 
\EE \big(\partial_\theta \Ss^{-1}_{\lambda,\sigma}(\theta)
\rho_\lambda (\Ss_{\lambda,\sigma}^{-1}(\theta))\big)
\;=\;
\rho_\lambda(\theta)
\;.
$$
Furthermore assuming that $\rho_\lambda$ is twice differentiable and
$\rho_\lambda=\rho+\Oo(\lambda^{\eta_{K+1}-\eta_K})$, this gives
the equation $\Ll^*\rho=0$ ({\it cf.} \cite{S} for further details).

\begin{theo}
\label{theo-FP}
The Fokker-Planck operator $\Ll^*$ has a unique groundstate $\rho$, 
which is non-negative, normalized and continuous except possibly in the
following cases:

\vspace{.1cm}

\noindent {\rm (I)} $p$ has a zero $\hat{\theta}$ of order $2$ {\rm (}and 
possibly another zero{\rm )} for which, moreover,
\begin{equation}
\label{eq-condI}
\partial_\theta\, p(\hat{\theta})
\;=\;
q(\hat{\theta})
\;=\; 0 \;,
\qquad
\partial_\theta^2\,p(\hat{\theta})
\;\geq\;
\partial_\theta \,q(\hat{\theta})
\;.
\end{equation}

\vspace{.1cm}

\noindent {\rm (II)} $p$ has a zero $\hat{\theta}$ of order $4$ for which,
moreover,
\begin{equation}
\label{eq-condII}
\EE(p_{K,\sigma}(\hat{\theta}))
\;=\;
0\;,
\qquad
\EE(\partial_\theta\;p_{K,\sigma}(\hat{\theta}))
\;\leq\;
0\;.
\end{equation}

\vspace{.1cm}

\noindent Furthermore, the groundstate is infinitely differentiable except
possibly at one point $\hat{\theta}$, at which $p$ has a zero of order $2$ and
\begin{equation}
\label{eq-condIII}
\partial_\theta\, p(\hat{\theta})
\;=\;
q(\hat{\theta})\;,
\qquad
\partial_\theta^2\,p(\hat{\theta})
\;<\;
\partial_\theta \,q(\hat{\theta})
\;.
\end{equation}
\end{theo}

Let us note that \eqref{eq-condI} is equivalent to
$$
\EE(\partial_\theta\;p_{1,\sigma}^2(\hat{\theta}))
\;=\;
4\,\EE(p_{K,\sigma}(\hat{\theta}))\;,
\qquad
\EE(\partial^2_\theta\;p_{1,\sigma}^2(\hat{\theta}))
\;>\;
4\,
\EE(\partial_\theta\;p_{K,\sigma}(\hat{\theta}))\;.
$$
Furthermore, let us point out that whenever there is a point $\hat{\theta}$
which is a zero of both $p$ and $q$, then the Dirac peak
$\delta_{\hat{\theta}}$ is a groundstate of $\Ll^*$ if considered as an
operator on distributions. This Dirac peak is dynamically stable under the
drift-diffusion described by $\Ll^*$ if, moreover,
$$
\partial^l\,\bigl((\partial_{\theta}p)\;-\;q\bigr)(\hat{\theta})
\;>\;0\;,
\qquad
l\,\in\,\{1,3\}
\;.
$$
This is precisely the case in (I) and (II). A corresponding dual result,
namely that the Birkhoff sums $I_N(f)$ are equal to  $f(\hat{\theta})$ to
lowest order,
will be proven in Theorem~\ref{theo-Birkhoff2} 
below. Let us also note that there may
be two such angles ({\it e.g.} if $\EE(p_{K,\sigma})=0$), in which case
the stable
groundstate is degenerate. Finally, we encourage the reader to draw the curves
of $p$ and $q$ in each of the cases of the proof below, and interprete $p$ as
a diffusive force, and $q$ as a drift. This allows to better 
understand the formal
proof, as well as the result itself.

\vspace{.2cm}

\noindent {\bf Proof} of Theorem~\ref{theo-FP}. The 
equation $\Ll^* \rho = 0$ can be integrated once. Hence one needs to solve
\begin{equation}
\label{eq-ODEred}
\bigl(p\,\partial_{\theta}
\;+\;
\tilde{q}\bigr)\,
\rho
\;=\;
C\;,
\qquad
\tilde{q}\;=\;(\partial_{\theta}p)\;-\;q\;,
\end{equation} 
\noindent where the real constant $C$ has to be chosen such 
that \eqref{eq-ODEred} admits a non-negative, $\pi$-periodic, $L^1$-normalized 
solution $\rho$. The equation \eqref{eq-ODEred} is locally precisely the
type of singular ODE studied in Appendix B, with coefficients which are
moreover real analytic. The only supplementary property of \eqref{eq-ODEred}
is that the inhomogeneity is constant. When writing out solutions, we use
similar notations as in the appendix, except that we
shall put tildes on $w$ and $W$
in order to distinguish them from the corresponding dual objects appearing in
the proof of Theorem~\ref{theo-Birkhoff} below, and
that we do not include the
constant inhomogeneity $C$ in the definition of $\tilde{W}$ below as done in
\eqref{eq-contsolution}. 

\vspace{.1cm}

If $p>0$ so that there is no
singularity and $\Ll^*$ is elliptic, the groundstate $\rho$
can readily be calculated as in \cite{S}. Set, for
some $\tilde{\theta}\in S^1_\pi$,
\begin{equation}
\label{eq-wW}
\tilde{w}(\theta)
\;=\;
\int^\theta_{\tilde{\theta}}
{\rm d}\xi\;
\frac{\tilde{q}(\xi)}{p(\xi)}
\;,
\qquad
\tilde{W}(\theta)
\;=\;
\int^\theta_{\tilde{\theta}}
{\rm d}\xi\;
\frac{e^{\tilde{w}(\xi)}}{p(\xi)}
\;.
\end{equation}
Then
\begin{equation}
\label{eq-regsol}
\rho
\;=\;
C_1\,e^{-\tilde{w}}\,\bigl(C_2\,\tilde{W}\,+\,1)
\;,
\qquad
C_2\;=\;
\frac{e^{\tilde{w}(\tilde{\theta}+\pi)}-1}{\tilde{W}(\tilde{\theta}+\pi)}
\;,
\qquad
C\;=\;C_1\,C_2\;,
\end{equation}
where one first determines the normalization constant $C_1$ which then also
fixes $C$ in \eqref{eq-ODEred} by the last identity. Note that
$\tilde{W}(\theta)>0$ for $\theta>\tilde{\theta}$ so that, in particular, 
$C_2$ is well-defined. Furthermore
$C_2=0$ if $w$ is $\pi$-periodic. The groundstate $\rho$ is clearly real
analytic in this case. 

\vspace{.1cm}

Now let $\hat{\theta}$ be the only zero of $p$. It is either of order $2$ or
$4$. If $\tilde{q}(\hat{\theta})\neq 0$, one is locally near $\hat{\theta}$ in
the situation of Proposition~\ref{prop-smoothsol}(iv) or (v). For each $C$ in
\eqref{eq-ODEred}, the smooth integrable solution is unique to one side of
$\hat{\theta}$. Continuing it cyclically around $S^1_\pi$ shows that it is
also unique on the other side of $\hat{\theta}$. 
For $\hat\theta \leq \theta < \hat\theta+\pi$,
this smooth solution is
given by 
\begin{equation}
\label{eq-solzero}
\rho
\;=\;
C\,e^{-\tilde{w}}\,\tilde{W}
\;,
\end{equation}
with $\tilde{w}$ and $\tilde{W}$ defined as in \eqref{eq-wW} using 
$\hat\theta<\tilde{\theta}<\hat{\theta}+\pi$ in the first equation and
$\tilde{\theta}=\hat\theta$ or $\tilde{\theta}=\hat{\theta}+\pi$ in 
the second one, such that the singularities cancel.
It only remains to choose $C\neq 0$ (and
hence the ODE \eqref{eq-ODEred} itself) such that $\rho$ is normalized.

\vspace{.1cm}

If $p$ has two zeros $\hat{\theta}_1$ and $\hat{\theta}_2$ of order $2$ for
which  $\tilde{q}(\hat{\theta}_1)\neq 0$ and  $\tilde{q}(\hat{\theta}_2)\neq
0$, one has to distinguish two cases. If the signs of
$\tilde{q}(\hat{\theta}_1)$ and $\tilde{q}(\hat{\theta}_2)$ are the same, then
one proceeds cyclically twice as in the case with one zero of $p$ which was
just treated. The solution is as in \eqref{eq-solzero} and $C\neq 0$. 
If on the other hand $0\leq \hat{\theta}_1<\hat{\theta}_2<\pi$ with
$\tilde{q}(\hat{\theta}_1)<0$ and $\tilde{q}(\hat{\theta}_2)>0$,
then one chooses $C=0$ and sets
\begin{equation}
\label{eq-splitted}
\rho(\theta)
\;=\;
\left\{
\begin{array}{ccc}
C_3\,e^{-\tilde{w}(\theta)}\, & & 
\hat{\theta}_1<\theta<\hat{\theta}_2\;,
\\
0 & & \mbox{otherwise ,}
\end{array}
\right.
\end{equation}
with $\tilde{w}$ defined as in \eqref{eq-wW} for some $\tilde{\theta}$ with
$\hat{\theta}_1<\tilde{\theta}<\hat{\theta}_2$ and some normalization constant
$C_3$. Note that this $\rho $ has no singularities because 
$\lim_{\theta\downarrow \hat{\theta}_1}\tilde{w}(\theta)=+\,\infty$ and
$\lim_{\theta\uparrow \hat{\theta}_2}\tilde{w}(\theta)=+\,\infty$. As shown in
Proposition~\ref{prop-smoothsol}(iv) and (v), $\rho$ is smooth at both
$\hat{\theta}_1$ and $\hat{\theta}_2$. 

\vspace{.1cm}

Now follow the singular cases where $p$ and $q$ have common zeros. The
situations (I) and (II) are of this type, but there will appear others
for which we then show as stated that there is a unique continuous groundstate.
As argued in Appendix B, there can then only be integrable solutions of 
\eqref{eq-ODEred} if the equation is homogeneous, namely $C=0$. One such
solution is $\rho=0$, but there may be others. First let $p$ have only one
zero $\hat{\theta}$ of order $2$ and suppose that
$\tilde{q}(\hat{\theta})=0$. Then $\hat{\theta}$ is a zero of order $2$ of 
$\tilde{q}$ if and only if $\partial_\theta\tilde{q}(\hat{\theta})=0$, that is 
$\EE(\partial^2_\theta\,p_{1,\sigma}^2(\hat{\theta}))=
4\,\EE(\partial_\theta\,p_{K,\sigma}(\hat{\theta}))$. In this situation, there
is a one-parameter family of smooth solutions according to 
Proposition~\ref{prop-smoothsol}(i). However, 
these solutions are typically not periodic and hence this is part of situation
(I). If $\hat{\theta}$ is a zero of order $1$ of 
$\tilde{q}$, then Proposition~\ref{prop-smoothsol}(ii) and (iii) can be used
for analyzing the local regularity. Case
(iii) applies if $\partial_\theta\tilde{q}(\hat{\theta})<0$.
Thus one  has a two-parameter family of continuous solutions near
$\hat{\theta}$; however, one parameter has to assure the periodicity of the
solution so that it remains continuous, while the second has to be chosen so
that $\rho$ is normalized. Hence the unique continuous groundstate is 
$\rho=C_4\,e^{-\tilde{w}}$ with an adequate normalization constant $C_4$. 
In the situation (I) when 
\eqref{eq-condI} holds with a strict inequality, one has  
$\partial_\theta\tilde{q}(\hat{\theta})>0$ so that 
Proposition~\ref{prop-smoothsol}(ii) implies that the solution is unique. As
$\rho=0$ is one solution, it is the only one. 

\vspace{.1cm}

Next let us consider the case where $p$ has two zeros 
$\hat{\theta}_1$ and $\hat{\theta}_2$ of order $2$, and that at least one of
them, say $\hat{\theta}_1$, is a zero of $\tilde{q}$. 
If $\partial_\theta\tilde{q}(\hat{\theta}_1)>0$
(namely situation (I) again), there is only the zero solution due to
Proposition~\ref{prop-smoothsol}(ii).

If $\partial_\theta\tilde{q}(\hat{\theta}_1)=0$ and 
$\tilde{q}(\hat{\theta}_2)\neq 0$, the singularity $\hat{\theta}_1$ 
can due
to Proposition~\ref{prop-smoothsol}(i) only be resolved 
by choosing $C=0$, but then
the zero $\hat{\theta}_2$ for which
$\tilde{q}(\hat{\theta}_2)\neq 0$ leads to a singularity. Thus there
is no continuous solution and this is included in situation (I). If 
$\partial_\theta\tilde{q}(\hat{\theta}_1)<0$ and say
$\tilde{q}(\hat{\theta}_2)> 0$, then one can appeal to 
Proposition~\ref{prop-smoothsol}(iii) at $\hat{\theta}_1$ and 
Proposition~\ref{prop-smoothsol}(iv) at $\hat{\theta}_2$, and construct the
groundstate exactly as in \eqref{eq-splitted}. However, $\rho$ might 
not be differentiable at  $\hat{\theta}_1$. If  
$\hat{\theta}_1$ and $\hat{\theta}_2$ are zeros of both $p$ and
$\tilde{q}$, then either
$\partial_\theta\tilde{q}(\hat{\theta}_1)>0$
or $\partial_\theta\tilde{q}(\hat{\theta}_2)>0$. Indeed,
$\tilde{q}=\frac{1}{2}\,\partial_\theta p-2\,\EE(p_{K,\sigma})$
and $\EE(p_{K,\sigma})$ is a trigonometric polynomial of degree $1$ and hence
can only compensate either 
$\partial_\theta p(\hat{\theta}_1)>0$ or
$\partial_\theta p(\hat{\theta}_2)>0$. Therefore this corresponds again to
situation (I) and there is no normalizable solution.

\vspace{.1cm}

In the last remaining case $\hat{\theta}$ is a zero of $p$ of order $4$ (hence
the only zero of $p$) and $\tilde{q}(\hat{\theta})=0$. If
$\partial_\theta\tilde{q}(\hat{\theta})<0$, which is equivalent to
$\EE(\partial_\theta\;p_{K,\sigma}(\hat{\theta}))>0$,
Proposition~\ref{prop-smoothsol}(vii) can be applied and the groundstate 
$\rho=C_5\,e^{-\tilde{w}}$ is smooth. If
$\partial_\theta\tilde{q}(\hat{\theta})>0$, then
Proposition~\ref{prop-smoothsol}(vi) implies that the zero solution is the
only solution. If $\partial_\theta\tilde{q}(\hat{\theta})=0$ and 
$\EE(p_{K,\sigma})=0$ identically, then 
$\tilde{q}=\frac{1}{2}\,\partial_\theta p$ has a zero at 
$\hat{\theta}$ of order $3$ and
$\partial^3_\theta\tilde{q}(\hat{\theta})>0$, so that 
Proposition~\ref{prop-smoothsol}(ii) implies that the zero solution is the
only solution. The same holds if 
$\partial_\theta\tilde{q}(\hat{\theta})=0$
and $\EE(\partial_\theta^2 p_{K,\sigma}(\hat{\theta}))\neq 0$ due to 
Proposition~\ref{prop-smoothsol}(iv) or (v), applied either to the left or
right of $\hat{\theta}$ pending on the sign of 
$\EE(\partial_\theta^2 p_{K,\sigma}(\hat{\theta}))$.
The last three cases are regrouped in situation (II).
\hfill $\Box$

\section{Birkhoff sums at second order anomalies}
\label{sec-2nd}

In this section we calculate the Birkhoff sums $I_N(f)$ perturbatively for many
second order anomalies. This involves the analysis of the operator $\Ll$ dual
to the Fokker-Planck operator. We will not present a treatment as exhaustive
as Theorem~\ref{theo-FP} for $\Ll^*$ and do not treat the singular cases where
$p$ and $q$ have common zeros, except for the case of
Theorem~\ref{theo-Birkhoff2} below, which is needed for the analysis of a band
edge. Comments on the remaining cases are
given at the end of the section.

\begin{theo}
\label{theo-Birkhoff}
Let $(T_{\lambda,\sigma})_{\sigma\in\Sigma}$ 
have an anomaly of second order.
Suppose $q(\hat\theta)\neq 0$ whenever
$\EE(p_{1,\sigma}^2(\hat\theta))=0$.
Furthermore let $\rho$ be the groundstate of the Fokker-Planck operator
given by {\rm Theorem~\ref{theo-FP}}. Then, for $f\in C^2(S^1_\pi)$ one has
%
$$
I_N(f)
\;=\;
\int_0^\pi {\rm d}\theta\;
f(\theta)\,\rho(\theta)
\,+\, \Oo\left(\lambda^{\tilde{\eta}-\eta_K},(\lambda^{\eta_K}N)^{-1} \right)
\;
$$
with $\tilde\eta=\min\{\eta_{\tilde{K}},\eta_1+\eta_2\}$ where
$\tilde{K}$ is the first uncentered term in {\rm \eqref{eq-expan}}
after $P_{\eta_K,\sigma}$.
If $\EE(p_{1,\sigma}^2)$ has no zero, it is enough to suppose $f\in
C^1(S^1_\pi)$. 
\end{theo}

\noindent {\bf Proof.}
As already argued in Section~\ref{sec-FP}, one can control the Birkhoff sums
using Proposition~\ref{prop-DGL} only if $f\in\,$Ker$(\Ll^*)^\perp$. Under the
hypothesis stated, the kernel of $\Ll^*$ is one-dimensional by
Theorem~\ref{theo-FP} and spanned by the groundstate $\rho$. 
Replacing $f$ by $f-\langle \rho|f\rangle$, we may now assume that
$\int_0^\pi {\rm d}\theta\,f(\theta)\,\rho(\theta)=0$ and have to show for
such $f$ that the Birkhoff sum $I_N(f)$ is of the order stated. This will
follow directly from Proposition~\ref{prop-DGL} once we have found $G\in
C^2(S^1_\pi)$ satisfying
\begin{equation}
\label{eq-G}
(p\,\partial_\theta\,+\,q)\,G\;=\;f
\;,
\qquad
\int_0^\pi
{\rm d}\theta\,G(\theta)
\;=\;
0\;.
\end{equation}
Indeed, the second identity allows to take an antiderivative $F$ of $G$
which then satisfies \eqref{eq-inhomODE}.

\vspace{.1cm}

First let us consider the case where $p$ has no zero and $C\neq 0$ in
\eqref{eq-ODEred}. Setting, for
some $\tilde{\theta}\in S^1_\pi$,
\begin{equation}
\label{eq-wW2}
{w}(\theta)
\;=\;
\int^\theta_{\tilde{\theta}}
{\rm d}\xi\;
\frac{{q}(\xi)}{p(\xi)}
\;,
\qquad
{W}(\theta)
\;=\;
\int^\theta_{\tilde{\theta}}
{\rm d}\xi\;
\frac{e^{{w}(\xi)}}{p(\xi)}\,f(\xi)
\;,
\end{equation}
the solution is given by
$$
G
\;=\;
e^{-{w}}\,\bigl({W}\,+\,c_1)
\;,
\qquad
c_1
\;=\;
\frac{e^{-w(\tilde\theta+\pi)}\,W(\tilde{\theta}+\pi)}
{1-e^{-{w}(\tilde{\theta}+\pi)}}
\;,
$$
where $e^{-{w}(\tilde{\theta}+\pi)}\neq 1$ as this happens if and only if 
$w$ and $\tilde w$ are $\pi$-periodic leading to $C=0$.
One readily checks that the first equation in \eqref{eq-G} is
satisfied and that $G$ is periodic, hence in $C^2(S^1_\pi)$ because
$f\in C^1(S^1_\pi)$. Furthermore, multiplying the first equation in 
\eqref{eq-G} by $\rho$ and integrating over $S^1_\pi$ shows that also the
second equation holds:
$$
0
\;=\;
\int\,\rho\,f
\;=\;
\int\,\rho\,\bigl(p\,\partial_\theta\,+\,q\bigr)G
\;=\;
-\,\int\,G\;\bigl(\partial_\theta\, p\,-\,q\bigr)\rho
\;=\;
-\,C\,
\int\,G(\theta)
\;.
$$

Next, if $p$ has no zeros and $C=0$ in \eqref{eq-ODEred}, then $C_2=0$ in 
\eqref{eq-regsol} and
\begin{equation}
\label{eq-dual}
\rho
\;=\;
C_1\,e^{-\tilde{w}}
\;=\;
C_1\,\frac{e^{{w}}}{p}
\;,
\qquad
\bigl(p\,\partial_\theta\,+\,q\bigr)\,e^{-{w}}
\;=\;0
\;,
\end{equation}
so that $W$ defined as in \eqref{eq-wW2} satisfies
$W(\tilde{\theta}+\pi)=W(\tilde{\theta})=0$ because $\int\rho f=0$. Hence
$$
G
\;=\;
e^{-w}\,(W\,+\,c_2)
$$
is a (periodic) function in $C^2(S^1_\pi)$ and, for adequate
choice of $c_2$, of vanishing integral. 

\vspace{.1cm}

Now let $p$ have one zero $\hat{\theta}$. The hypothesis implies
$q(\hat{\theta})\neq 0$. According to 
Proposition~\ref{prop-smoothsol}(iv) or (v), 
the first equation in \eqref{eq-G} has a $C^2$-solution
in a neighborhood of
$\hat{\theta}$, which is unique to one side of $\hat{\theta}$. Continuing this
solution around $S^1_\pi$ one deduces that the solution is unique, and
actually given by $G=e^{-w}W$. Due to the argument in the proof of 
Theorem~\ref{theo-FP}, $C\neq 0$ in \eqref{eq-ODEred} in this case, so that by
the same argument as above the second equation in \eqref{eq-G} is
satisfied. The case with two zeros $\hat{\theta}_1$ and $\hat{\theta}_2$
of $p$ for which $q(\hat{\theta}_1)$ and $q(\hat{\theta}_2)$ have the same
sign is treated similarly ({\it cf.} the proof of Theorem~\ref{theo-FP}). 

\vspace{.1cm}

Finally we deal with the case of two zeros $0\leq
\hat{\theta}_1<\hat{\theta}_2<\pi$ 
of $p$ for which $q(\hat{\theta}_1)>0$ and $q(\hat{\theta}_2)<0$. 
By \eqref{eq-splitted}, the support of the groundstate is 
$[\hat{\theta}_1,\hat{\theta}_2]$ and hence not all of $S^1_\pi$. 
For $\theta\in[\hat{\theta}_1,\hat{\theta}_2]$, we choose
$\hat{\theta}_1<\tilde\theta<\hat{\theta}_2$ in the first equation
of \eqref{eq-wW2} and
$\tilde\theta=\hat{\theta}_1$ in the second equation of
\eqref{eq-wW2}. Then $G=e^{-w}\,W$ is a solution 
in $(\hat{\theta}_1,\hat{\theta}_2)$. Due to
Propostion~\ref{prop-smoothsol}(iv) this is the unique solution that
can be smoothly extended to the left of $\hat{\theta}_1$. 
Furthermore \eqref{eq-dual} and $\int\rho f=0$
imply $W(\hat{\theta}_2)=0$, so that this solution can also be 
smoothly extended through $\hat{\theta}_2$ by
Propostion~\ref{prop-smoothsol}(v). On both sides one has one free
parameter. One is chosen such that $G$ is 
$\pi$-periodic and the remaining one such that the 
second equation of \eqref{eq-G} is also satisfied. This reflects that
$G_0= e^{-w}\chi_{[\hat{\theta}_2,\hat{\theta}_1]}$ is a smooth
positive solution of the homogeneous equation 
$(p\,\partial_\theta+q)G_0=0$. 
\hfill $\Box$

\vspace{.2cm}

The remaining singular cases for which Theorem~\ref{theo-FP}
guarantees existence 
of a unique groundstate cannot be treated by the techniques of 
Theorem~\ref{theo-Birkhoff}
without further hypothesis on $f$,
which seem somewhat unnatural, but reflect the delicate dynamical behavior at
such points 
({\it e.g.} one needs that $f$ and its derivative have a particular
behavior near the common zero of $p$ and $q$). 
The next theorem deals with an anomaly corresponding to situation (II) of 
Theorem~\ref{theo-FP}, albeit with a strict inequality in \eqref{eq-condII}.

\begin{theo}
\label{theo-Birkhoff2}
Let $(T_{\lambda,\sigma})_{\sigma\in\Sigma}$ 
have an anomaly of second order. 
Suppose that $\hat{\theta}$ is a zero of
order $4$ of $\,\EE(p_{1,\sigma}^2)$ and a zero of order $1$ of 
$\,\EE(p_{K,\sigma})$ and that, moreover,
$\EE(\partial_\theta p_{K,\sigma}(\hat\theta))<0$.
Then for $f\in C^2(S^1_\pi)$ 
$$
I_N(f)
\;=\;
f(\hat{\theta})
\,+\, \Oo\left(\lambda^{\tilde{\eta}-\eta_K},(\lambda^{\eta_K}N)^{-1} \right)
\;
$$
with $\tilde\eta=\min\{\eta_{\tilde{K}},\eta_1+\eta_2\}$ where
$\tilde{K}$ is the first uncentered term in {\rm \eqref{eq-expan}}
after $P_{\eta_K,\sigma}$.
\end{theo}

\noindent {\bf Proof.} The procedure is exactly as in the proof of
Theorem~\ref{theo-Birkhoff}. Hence we search for a solution of
\eqref{eq-G} for a function $f$ satisfying $f(\hat{\theta})=0$. Close to 
$\hat{\theta}$ the equation $(p\,\partial_\theta+q)G=f$ has a two-parameter
family of solutions due to 
Proposition~\ref{prop-smoothsol}(vii). One parameter is fixed by requiring $G$
to be periodic. The other reflects that one can, by the same procedure, find a
non-vanishing solution $G_0$ to 
$(p\,\partial_\theta+q)G_0=0$. As $G_0$ has definite sign, one can use it to
normalize $G$ so that also the second equation in 
\eqref{eq-G} is satisfied.
\hfill $\Box$

\vspace{.2cm}

As to extensions of Theorem~\ref{theo-Birkhoff2}, the case with an equality in
the second equation of \eqref{eq-condII} cannot be treated by the presented
technique because $(p\,\partial_\theta+q)G_0=0$ has only the trivial solution
by Proposition~\ref{prop-smoothsol}(iv) or (v). The case with one zero of
order $2$ corresponding to situation (I) in Theorem~\ref{theo-FP} can be
treated similarly as long as 
$3\,\EE(\partial^2_\theta\;p_{1,\sigma}^2(\hat{\theta}))<
4\,|\EE(\partial_\theta\;p_{K,\sigma}(\hat{\theta}))|$, a condition which
originates  in Proposition~\ref{prop-smoothsol}(iii).

\section{Application to a band edge}
\label{sec-bandedge}

This section contains the proof of Theorem~\ref{theo-scaling}. Hence let
$T_{\lambda,\sigma}=N_{\lambda,\frac{\eta}{2}}
N\,\Tt^{E_b+\epsilon\lambda^\eta}_{\lambda,\sigma}\,N^{-1}
N_{\lambda,\frac{\eta}{2}}^{-1}$ be the anomaly given in
\eqref{eq-bandanom} or \eqref{eq-bandanom2}. Then 
\eqref{eq-links} allows to calculate the Lyapunov exponent and the IDS, so
that one may focus on the calculation of $\gamma(\lambda)$ and $\Rr(\lambda)$
based on \eqref{eq-lyap}, \eqref{eq-rot} and
the expansions given in Lemma~\ref{lem-coefficients}.

\vspace{.2cm}

Let us begin with the elliptic first order regime, hence \eqref{eq-bandanom}. 
The adequate basis change
assuring \eqref{eq-ellipchange} is
$$
M
\;=\;
\left(
\begin{matrix} 
1 & 0 \\
0 & -\,|\epsilon\,x|^{-\frac{1}{2}} 
\end{matrix}
\right)
\;.
$$
Then $(MN_{\lambda,\delta}
N)\Tt^{E_b+\epsilon\lambda^\eta}_{\lambda,\sigma}(M
N_{\lambda,\delta}N)^{-1}$ is equal to
$$
\exp
\left(
\,
\lambda^{\frac{\eta}{2}}
|\epsilon\,x|^{\frac{1}{2}} 
\left(
\begin{matrix} 
0 & -1 \\
1 & 0
\end{matrix}
\right)
\,-\,
\lambda^{1-\frac{\eta}{2}}
\frac{x_\sigma}{|\epsilon\,x|^{\frac{1}{2}} }
\left(
\begin{matrix} 
0 & 0 \\
1 & 0
\end{matrix}
\right)
\,+\,
\lambda^\eta\,P_\eta
\,+\,
\lambda\,P_{1,\sigma}
\,+\,
\Oo(\lambda^{\frac{3\eta}{2}},\lambda^{1+\frac{\eta}{2}})
\,
\right)
\;.
$$
Here $\EE(P_{1-\frac{\eta}{2},\sigma})=0$, and also $\EE(P_{1,\sigma})=0$.
With the notations of Lemma~\ref{lem-coefficients}, this gives
$\beta_{\frac{\eta}{2}}=0$ and
$\beta_{1-\frac{\eta}{2},\sigma}=
\imath\,x_\sigma|\epsilon\,x|^{-\frac{1}{2}}/2$. Thus
Proposition~\ref{prop-ellipt} implies 
$I(e^{2j\imath \theta})=\Oo(\lambda^{\frac{\eta}{2}},\lambda^{2-\frac{3\eta}{2}})$ 
for $j=1,2$. 
Hence the first 4 terms with powers $\lambda^{\eta_k}$ in \eqref{eq-expan1}
either vanish or are of higher order, the same holds for the mixed terms
$\lambda^{\eta_{k_1}+\eta_{k_2}}$, the first of the terms with
$\lambda^{2\eta_k}$ is also of higher order, but the second one gives the
leading order contribution:
$$
\gamma(\lambda)
\;=\;
\lambda^{2-\eta}\;
\frac{\EE(x^2_\sigma)}{8\,|\epsilon\,x|}
\;+\;
\Oo(\lambda^{\frac{3\eta}{2}},\lambda^{4-\frac{5\eta}{2}}
)
\;,
$$
which together with \eqref{eq-links} establishes the first claim. Similarly
one verifies $\alpha_{\frac{\eta}{2}}=\imath |\epsilon\,x|^{\frac{1}{2}}$ so
that by \eqref{eq-expan2} one finds  
$\Rr(\lambda)=\lambda^{\frac{\eta}{2}}\,|\epsilon\,x|^{\frac{1}{2}}
+\Oo(\lambda^{\eta},\lambda^{2-\eta})$.

\vspace{.2cm}

In the hyperbolic regime, the basis change
leading to \eqref{eq-hypchange} is
$$
M
\;=\;
\left(
\begin{matrix} 
1 & -\,|\epsilon\,x|^{-\frac{1}{2}}  \\
1 & |\epsilon\,x|^{-\frac{1}{2}} 
\end{matrix}
\right)
\;.
$$
Then $(MN_{\lambda,\frac{\eta}{2}}
N)\Tt^{E_b+\epsilon\lambda^\eta}_{\lambda,\sigma}(M
N_{\lambda,\frac{\eta}{2}}N)^{-1}$ calculated from \eqref{eq-bandanom} is equal to
$$
\exp
\left(
\,
\lambda^{\frac{\eta}{2}}
|\epsilon\,x|^{\frac{1}{2}} 
\left(
\begin{matrix} 
-1 & 0 \\
0 & 1
\end{matrix}
\right)
\,+\,
\lambda^{1-\frac{\eta}{2}}
\frac{x_\sigma}{2\,|\epsilon\,x|^{\frac{1}{2}} }
\left(
\begin{matrix} 
-1 & -1 \\
1 & 1
\end{matrix}
\right)
\,+\,
\lambda^\eta\,P_\eta
\,+\,
\lambda\,P_{1,\sigma}
\,+\,
\Oo(\lambda^{\frac{3\eta}{2}},\lambda^{1+\frac{\eta}{2}})
\,
\right)
\;.
$$
Now let us apply a further basis change $N_{\lambda,\delta}$ as described in 
\eqref{eq-hyptransform}, namely in the present situation
$\eta_\Kk=\frac{\eta}{2}$, $\chi=1-\frac{\eta}{2}$ and $\delta =1-
\frac{3\eta}{4}$. Due to \eqref{eq-rescale} the transformed anomaly then
becomes of second order: 
$$
\exp
\left(
\,
\lambda^{\frac{\eta}{4}}
|4\,\epsilon\,x|^{-\frac{1}{2}} 
\left(
\begin{matrix} 
0 & 0 \\
x_\sigma & 0
\end{matrix}
\right)
\,+\,
\lambda^{\frac{\eta}{2}}
|\epsilon\,x|^{\frac{1}{2}} 
\left(
\begin{matrix} 
-1 & 0 \\
0 & 1
\end{matrix}
\right)
\,+\,
\lambda^{1-\frac{\eta}{2}}
\,P'_{1-\frac{\eta}{2},\sigma}
\,+\,
\lambda^{2-\frac{5\eta}{4}}
\,P'_{2-\frac{5\eta}{4},\sigma}
\,+\,
\Oo
\,
\right)
\;,
$$
where $\Oo=\Oo(\lambda^{\frac{3\eta}{4}},\lambda^{\frac{7\eta}{4}-1})$. Here
$P'_{1-\frac{\eta}{2},\sigma}$ and $P'_{2-\frac{5\eta}{4},\sigma}$ are
possibly of lower order than the second term 
(but not as the first one as $\eta<\frac{4}{3}$), but they are both centered
so that they do not enter into the asymptotics ({\it cf.}
Definition~\ref{def-critical}). The obtained second anomaly can now be dealt
with by Theorem~\ref{theo-Birkhoff2}, implying that $I(f)=f(\frac{\pi}{2})+
\Oo(\lambda^{\frac{5\eta}{4}-1},\lambda^{1-\frac{3\eta}{4}})$. 
Furthermore, one calculates
$\beta_{\frac{\eta}{4},\sigma}=
-\imath\,x_\sigma|\epsilon\,x|^{-\frac{1}{2}}/4$,
$\gamma_{\frac{\eta}{4},\sigma}=
x_\sigma^2|\epsilon\,x|^{-1}/8$ and 
$\beta_{\frac{\eta}{2}}=-|\epsilon\,x|^{\frac{1}{2}}$.
Replacing into \eqref{eq-expan1} one realizes that the factor of
$\lambda^{2\frac{\eta}{4}}$ vanishes so that
$$
\gamma(\lambda)
\;=\;
\lambda^{\frac{\eta}{2}}\;|\epsilon\,x|^{\frac{1}{2}}
\;+\;
\Oo(\lambda^{\frac{7\eta}{4}-1},\lambda^{1-\frac{\eta}{4}})
\;.
$$
Concerning the rotation number,
one has $\alpha_{\frac{\eta}{2}}=0$ and 
$\alpha_{\frac{\eta}{4},\sigma}=
\imath\,x_\sigma|\epsilon\,x|^{-\frac{1}{2}}/4$. The latter implies that the
contribution $\lambda^{2\frac{\eta}{4}}$ vanishes in \eqref{eq-expan2} due to
the $\Im m$. Thus
$\Rr(\lambda)=\Oo(
\lambda^{\frac{7\eta}{4}-1},\lambda^{1-\frac{\eta}{4}})$.
Note that the condition $\frac{7\eta}{4}-1>\frac{\eta}{2}$ is equivalent to 
$\eta>\frac{4}{5}$. 

\vspace{.2cm}

Finally let us deal with the parabolic case for which \eqref{eq-bandanom2} is
directly a second order anomaly. The polynomials  \eqref{eq-polynomials} are
explicitly given by
$$
p(\theta)
\;=\;
\EE(x_\sigma^2)\,\cos^4(\theta)\;,
\qquad
q(\theta)
\;=\;
\epsilon\, x\,-\,1\,+\,(\epsilon\,x\,+\,1)\,\cos(2\theta)\,-2\,
\EE(x_\sigma^2)\,\cos^3(\theta)\,\sin(\theta)
\;.
$$
Because $q(\frac{\pi}{2})=-2\neq 0$, Theorem~\ref{theo-FP} guarantees
the existence of a unique normalized groundstate $\rho\in
C^\infty(S^1_\pi)$. As $\alpha_{\frac{1}{3},\sigma}=\imath\,x_\sigma/2$, 
$\beta_{\frac{1}{3},\sigma}=-\imath\,x_\sigma/2$ and 
$\gamma_{\frac{1}{3},\sigma}=x^2_\sigma/2$, one deduces from 
Lemma~\ref{lem-coefficients} and Theorem~\ref{theo-Birkhoff} that up to errors
of order $\Oo(\lambda)$
$$
\gamma(\lambda)
\;=\;
\lambda^{\frac{2}{3}}
\left(
\frac{1}{2}\,(\epsilon\,x\,+\,1)\,
\int {\rm d}\theta\;\rho(\theta)\,\sin(2\,\theta)
\;+\;
\frac{\EE(x_\sigma^2)}{8}\,
\int {\rm d}\theta\;\rho(\theta)\,(1\,+\,2\cos(2\,\theta)
\,+\,\cos(4\theta))
\right)
,
$$
and, furthermore that $\Rr(\lambda)$ is, also up to errors of order
$\Oo(\lambda)$, equal to
$$
\frac{\lambda^{\frac{2}{3}}}{2\,\pi}\,
\left(
\epsilon\,x\,-\,1\,
+
\,
(\epsilon\,x\,+\,1)
\int {\rm d}\theta\;\rho(\theta)\,\cos(2\,\theta)
\;-\;
\frac{\EE(x_\sigma^2)}{4}\,
\int {\rm d}\theta\;\rho(\theta)\,(2\,\sin(2\,\theta)
\,+\,\sin(4\theta))
\right)
\;.
$$

Let us remark that this formula is 
{\bf not} the same as equation (35) of \cite{DG}, even though the invariant
measure of \cite{DG} coincides with the above. In fact, their expansion
of the expression $\log(1+\lambda^{2/3} t)\approx\lambda^{2/3} t$
is erroneous for large $t$.

\section*{Appendix A: Transfer matrix at a band edge}
\label{sec-appendix}

Let us consider a real analytic family $E\in\RR\mapsto \Tt^E\in{\rm
SL}(2,\RR)$ with
$$
\Tr(\Tt^0)\;=\;2\;,
\qquad
\partial_E \Tr(\Tt^E)|_{E=0}
\;\neq\;
0
\;.
$$
If $\Tt^E$ is the transfer matrix of a periodic Jacobi matrix, this
corresponds to a band edge at $E=0$. We
show that necessarily the eigenvalue $1$ of $\Tt^0$ has 
geometric multiplicity $1$, so that the Jordan form of $\Tt^0$ is
non-trivial. A similar statement holds if $\Tr(\Tt^0)=-2$.
For this purpose, let us choose the notation
$$
\Tt^E
\;=\;
\left(\begin{matrix} 
a+A\,E & b+B\,E  \\ c+C\,E & d+D\,E 
\end{matrix} \right)
\;+\;
\Oo(E^2)
\;.
$$
Then one deduces $ad-bc=1$, $a+d=2$ and $A+D\neq 0$. Furthermore
the order $E$ of $\det(\Tt^E)=1$ implies that $aD+dA-cB-bC=0$. We need
to show that $\Tt^0\neq {\bf 1}$. If $a\neq 1$, this is true. Hence
suppose $a=1$ so that also $d=1$. Then  $ad-bc=1$ implies that either
$b=0$ or $c=0$.  Hence $aD+dA-cB-bC=0$ implies that either $A+D-cB=0$
or $A+D-bC=0$. Because $A+D\neq 0$ it follows that either $c\neq 0$ or
$b\neq 0$. 

\section*{Appendix B: Inhomogeneous singular first order ODE}
\label{sec-appendix2}

This appendix recollects some results about the solutions of the
inhomogeneous first order ordinary differential equation
(ODE) for a function $y=y(x)$ on the interval $U=(a,b)$
\begin{equation}
\label{eq-singODE}
p\,y'\,
+\,
q\,y
\;=\;r\;,
\end{equation}
where $p,q,r\in C^m(U)$ and $m\in \NN\cup\{\infty\}$. In particular,
we are interested in smooth solutions of
the singular case where $p$, $q$ and $r$ have
zeros at $\hat{x}\in U$ of order $0\leq l_p,l_q< m$ and $l_r$
respectively. We suppose that $p$ and $q$ have no further zeros in $U$. 
If $l_r<\min\{l_p,l_q\}$, then there cannot be any solution
$y\in C^1(U)$ because then
$$
\frac{p}{r}\,y'\,
+\,
\frac{q}{r}\,y
\;=\;1\;
$$
leads to the contradiction $0=1$ in the limit $x\to\hat{x}$. Hence let us
suppose that  $l_r\geq \min\{l_p,l_q\}$. In the
regular case $l_p\leq l_q$, the homogeneous equation 
$py'+qy=0$ has a one parameter
family of solutions $e^{-w}$ where $w(x)=\int^x \frac{q}{p} \in
C^{m-l_p+1}(U)$ is any antiderivative. Then a one-parameter family of 
solutions of \eqref{eq-singODE} is obtained by the method of
variation of constants:
\begin{equation}
\label{eq-regODE}
y(x)
\;=\;
e^{-w(x)}
\int^x{\rm d}s \;e^{w(s)}\;\frac{r(s)}{p(s)}\;,
\end{equation}
where the integral is well-defined because $l_r\geq l_p$. In the
singular case, one has $l_q<l_p$ and $l_q\leq l_r$ and the zero of $p$
effectively decouples the intervals $(a,\hat{x})$ and $(\hat{x},b)$. 
Any antiderivative 
$w(x)=\int^x \frac{q}{p}$ diverges as $x\to\hat{x}$. Nevertheless, we
shall show below that these divergences may cancel out in 
\eqref{eq-regODE}. Let us note right away though that any solution
$y\in C^1(U)$ has to satisfy
\begin{equation}
\label{eq-boundarycond}
y(\hat{x})
\;=\;
\lim_{x\to\hat{x}}
\;\frac{r(x)}{q(x)}
\;.
\end{equation}
This is a boundary condition for both intervals
$(a,\hat{x})$ and $(\hat{x},b)$. The question is whether there are
differentiable solutions (at the point $\hat{x}$, in particular) 
and whether such a solution is unique, or there are one-parameter or
two-parameter families of such solutions.

\begin{proposi}
\label{prop-smoothsol} 
Let $l_r\geq \min\{l_q,l_p\}$ and set $l=l_p-l_q$.
Then the ODE  {\rm \eqref{eq-singODE}}
has continuous solutions $y$ 
with a number of free parameters as described in the following:

\vspace{.1cm}

\noindent {\rm (i)} If $l\leq 0$, there is a one-parameter family of
solutions $y\in C^{m-l_p+1}(U)$.

\vspace{.1cm}

\noindent {\rm (ii)} If $l=1$ and $\partial\frac{p}{q}(\hat{x})>0$, the
solution $y\in C^{m-l_q}(U)$ is unique.

\vspace{.1cm}

\noindent {\rm (iii)} If $l=1$ and $\partial\frac{p}{q}(\hat{x})<0$,
then there is a two-parameter family of solutions $y\in C^n(U)$ as

long as $n\leq m-l_q$ and $n<|\partial \frac{p}{q}(\hat{x})|^{-1}$.

\vspace{.1cm}

\noindent {\rm (iv)} If $l\geq 2$ is even and
$\partial^l\frac{p}{q}(\hat{x})>0$, the 
solution $y\in C^{m-l_q}(U)$ is unique for
$x>\hat{x}$, but there
is one free parameter 
for $x<\hat{x}$. 

\vspace{.1cm}

\noindent {\rm (v)} If $l\geq 2$ is even and
$\partial^l\frac{p}{q}(\hat{x})<0$, the 
solution $y\in C^{m-l_q}(U)$ is unique for
$x<\hat{x}$, but 
there 
is 
one free parameter 
for $x>\hat{x}$. 

\vspace{.1cm}

\noindent {\rm (vi)} If $l\geq 3$ is odd and 
$\partial^l\frac{p}{q}(\hat{x})>0$, the 
solution $y\in C^{m-l_q}(U)$ is unique.

\vspace{.1cm}

\noindent {\rm (vii)} If $l\geq 3$ is odd and 
$\partial^l\frac{p}{q}(\hat{x})<0$, there is a two-parameter family of 
solutions $y\in C^{m-l_q}(U)$.

\vspace{.2cm}

\noindent Any other solution $y$ has a non-integrable singularity at
$\hat{x}$. 

\end{proposi} 

\noindent {\bf Proof.} In case (i) the solutions are given in
\eqref{eq-regODE}. Next let us consider the case (iv). Dividing 
\eqref{eq-singODE} by $q$ gives
$\frac{p}{q}\,y'+y=\frac{r}{q}$. Hence it is sufficient to consider the
case $p\,y'+y=r$ with $p,r\in C^{m-l_q}(U)$ and $l_p=l$. Let $w$ be any
antiderivative of $\frac{1}{p}$ on $U\setminus \{\hat{x}\}$.  Because
of  $\partial^l p(\hat{x})>0$ and $l$ is even, one has 
$\lim_{x\uparrow \hat{x}}w(x)=+\infty$ and 
$\lim_{x\downarrow \hat{x}}w(x)=-\infty$. Moreover, the growth of 
$|w(x)|$ is at least as $|x-\hat{x}|^{-1}$ because $l\geq 2$. 
Thus for all $n\in\NN$
\begin{equation}
\label{eq-asympt}
\lim_{x\uparrow\hat{x}}
\;
p^n(x)
\,e^{w(x)}
\;=\;
+\;\infty
\;,
\qquad
\lim_{x\downarrow\hat{x}}
\;
\frac{e^{w(x)}}{p(x)}
\;=\;
0
\;.
\end{equation}
Now $e^{-w}$ is a solution of the homogeneous equation
$py'+y=0$. Because $\lim_{x\downarrow \hat{x}}w(x)=-\infty$, the only
possible continuous solution of $p\,y'+y=r$
for $x>\hat{x}$ is 
\begin{equation}
\label{eq-contsolution}
y(x)
\;=\;
e^{-w(x)}\,W(x)\;,
\qquad
W(x)
\;=\;
\int^x_{\hat{x}}\;{\rm d}s\;
e^{w(s)}\;\frac{r(s)}{p(s)}
\;,
\end{equation}
where the last integral exists due to \eqref{eq-asympt}.
For $x<\hat{x}$, let $W$ be any antiderivative of $e^w\frac{r}{p}$ and
also set $y=e^{-w}W$. Hence $y$ is now a solution of 
$p\,y'+y=r$ for $x\neq \hat{x}$. It remains to be shown that $y\in
C^{m-l_q}(U)$. First of all, $y$ is continuous because
$$ 
\lim_{x\to \hat x}\,y(x)
\;=\;
\lim_{x \to \hat x} 
\frac{W(x)}{e^{w(x)}} 
\;=\; 
\lim_{x \to \hat x} \frac{e^{w(x)}\;\frac{r(x)}{p(x)}}{e^{w(x)}
\frac{1}{p(x)}} 
\;=\; 
r(\hat x)
\;=\; 
y(\hat x)
\;, 
$$ 

\noindent the latter by \eqref{eq-boundarycond}.  
In the second equality, de l'Hospital's rule could 
be applied because for $x\downarrow \hat x$ 
the numerator and denominator both converge 
to 0 by \eqref{eq-asympt} and \eqref{eq-contsolution}, while for 
$x\uparrow \hat x$ as the denominator 
converges to $\infty$ by \eqref{eq-asympt}.

\vspace{.2cm}

Next follows an inductive argument in order to check the 
continuity of the higher derivatives of $y$ at $\hat{x}$. Let us set:
$$
q_n\;=\;n\,p'\,+\,1\;,
\qquad
r_n\;=\;r'_{n-1}\,-\,q_{n-1}'\,y^{(n-1)}
\;,
\qquad
r_0\;=\;r\;.
$$
Note that $q_n$ has no zero at $\hat{x}$. One can check by induction 
\begin{equation}
\label{eq-identity1}
p\,y^{(n+1)}\,+\,q_n\,y^{(n)}
\;=\;
r_n
\;.
\end{equation}
Let $w_n=w+n\,\log(|p|)$. Now $e^{-w_n}$ is a solution of the homogeneous equation
$p\,y^{(n+1)}+q_n\, y^{(n)}=0$ and satisfies due to \eqref{eq-asympt}
\begin{equation}
\label{eq-asympt2}
\lim_{x\uparrow\hat{x}}
\;
p(x)
\,e^{w_n(x)}
\;=\;
+\;\infty
\;,
\qquad
\lim_{x\downarrow\hat{x}}
\;
e^{w_n(x)}
\;=\;
0
\;.
\end{equation}
Now let $y^{(n)}$, $n<m-l_g$, be continuous by induction
hypothesis. Then $r_n\in C^1(U)$. Set $W_n=y^{(n)}\,e^{w_n}$. Due to 
\eqref{eq-asympt2}, one has 
$\lim_{x\downarrow \hat{x}}W_n(x)=0$. Using the identities
$$
y^{(n+1)} 
\;=\; 
\frac{r_n\,e^{w_n}-q_n W_n}{e^{w_n} p}\;,
\qquad
q_n\,W_n'
\;=\;
r_n\,w_n'\,e^{w_n} 
$$
which follow from \eqref{eq-identity1} and $w_n'=\frac{q_n}{p}$,
one obtains
\begin{eqnarray*} 
\lim_{x \to \hat x} 
y^{(n+1)}(x) 
& = &
\lim_{x \to \hat x} \frac{-\,q_n(x)\, W_n(x) \,+\,
r_n(x) \,e^{w_n(x)}}{e^{w_n(x)} \,p(x)} 
\\ 
& = & 
\lim_{x \to \hat x} 
\frac{-\,q_n'(x)\,W_n(x)\,+\,r_n'(x)\,e^{w_n(x)}}{
e^{w_n(x)}(p'(x)\,+\,q_n(x))} 
\;=\; 
\frac{-q_n'(\hat x)\,y^{(n)}(\hat x)\,+\,r_n'(\hat x)}{q_{n+1}(\hat
x)}
\;,
\end{eqnarray*} 
where de l'Hospital's rule can be used for $x\downarrow \hat x$, 
because the numerator and denominator converge 
both to 0 according to (\ref{eq-asympt2}) and 
for $x \uparrow \hat x$ as the denominator converges 
to $\pm \infty$. Therefore $y^{(n)}$ is 
continuously differentiable in $\hat x$.

\vspace{.2cm}

The case (v) is identical, once the sides $x>\hat{x}$ and
$x<\hat{x}$ are exchanged in the above argument. In 
case (vi), one has $\lim_{x\to \hat{x}}w(x)=-\,\infty$ (namely, for
both $x\downarrow\hat{x}$ and $x\uparrow\hat{x}$),
so that
one has to construct the solution as in 
\eqref{eq-contsolution} both for $x>\hat{x}$ and $x<\hat{x}$; hence
the solution is unique. In case (vii), one has 
$\lim_{x\to \hat{x}}w(x)=+\,\infty$ and the homogeneous solutions on
both sides vanish at $\hat{x}$ faster than any power, so that one has a
two-parameter family of solutions.

\vspace{.2cm}

Finally let us consider $l=1$ and first the case (ii). Hence
$\lim_{x\to \hat{x}}w(x)=-\,\infty$ so that 
$\lim_{x\to \hat{x}}e^{w(x)}=0$. 
Now the second limit in (\ref{eq-asympt}) may diverge to $+\,\infty$,
but one readily checks that $\frac{e^w}{p}$ is integrable so that $W$
can be defined as in \eqref{eq-contsolution} for  
$x>\hat{x}$ and $x<\hat{x}$. Now the continuity is proven as above,
and higher derivatives can also be treated as above because
$$
\lim_{x\to\hat{x}}
\;
\,e^{w_n(x)}
\;=\;0\;,
$$
so that both nominator and numerator of $y^{(n+1)}$ converge to $0$
and de l'Hopital's rule may be applied as before. 

\vspace{.2cm}

In case (iii), $\lim_{x\to \hat{x}}w(x)=\infty$ so that 
$\lim_{x\to \hat{x}}e^{-w(x)}=0$. Furthermore $\frac{e^w}{p}$ is never
integrable and the two-parameter family of solutions $y$ is defined as
in \eqref{eq-regODE}. The main point to note is that $w_n=(p'(\hat{x})^{-1}
+n)\log (|x-\hat{x}|)+\Oo(1)$, hence
$$
\lim_{x\uparrow\hat{x}}
\;
\,p(x)\,e^{w_{n-1}(x)}
\;=\;\infty\;,
\qquad
\lim_{x\downarrow\hat{x}}
\;
\,p(x)\,e^{w_{n-1}(x)}
\;=\;-\,\infty\;,
$$
under the hypothesis stated in (iii). Thus de l'Hopital's rule 
can be invoked to calculate $\lim_{x\to \hat{x}}y^{(n)}$ as above (with the
index  $n$ shifted by $1$). Let us remark that in case
$|p'(\hat{x})|^{-1}\in\NN$, one can prove better differentiability, but only
for a one-parameter family of solutions 
(just as for the ODE $x\,y'=N\,y$, $N\in\NN$).  

\vspace{.2cm}

The singularity of all other solutions is at least of the type
$|x-\hat{x}|^{-1}$ in cases (ii) to (vii). 
\hfill $\Box$


\end{document}